\documentclass[a4paper,12pt]{article}

\pdfoutput=1
\usepackage[paper=letterpaper,margin=1.25in]{geometry}

\usepackage{hyperref}
\usepackage{graphicx}
\usepackage{dcolumn}
\usepackage{float}
\usepackage{longtable}
\usepackage{amsmath}
\usepackage{amssymb}
\usepackage{caption}
\usepackage{cite}

\usepackage{xcolor}
\definecolor{light-gray}{gray}{0.75}

\newlength{\xtrawidth}
\newlength{\xtraheight}
\include{graphix}

\begin{document}

\parskip 0.05in
\begin{flushright}
{\normalsize KIAS-P13050}\\
\end{flushright}

\begin{centering}
\vspace*{1.2cm}
{\Large \bf  Heterotic Model Building: 16 Special Manifolds}

\vspace{1cm}

{\large \bf Yang-Hui He$^{1,}$}\footnote{hey@maths.ox.ac.uk},
{\large \bf Seung-Joo Lee$^{2,}$}\footnote{s.lee@kias.re.kr},
{\large \bf Andre Lukas$^{3,}$}\footnote{lukas@physics.ox.ac.uk},
{\large \bf Chuang Sun$^{3,}$}\footnote{chuang.sun@physics.ox.ac.uk}

{\small
\vspace*{.5cm}
${}^1$
Department of Mathematics, City University, London, EC1V 0HB, UK; \\
School of Physics, NanKai University, Tianjin, 300071, P.R.~China; \\
Merton College, University of Oxford, OX14JD, UK
\\[0.3cm]
${}^{2}$ School of Physics, Korea Institute for Advanced Study, Seoul 130-722, Korea
\\[0.3cm]
${}^{3}$ 
Rudolf Peierls Centre for Theoretical Physics, University of Oxford\\
1 Keble Road, Oxford OX1 3NP, UK\\[1cm]
}

\begin{abstract}\noindent
We study heterotic model building on 16 specific Calabi-Yau manifolds constructed as hypersurfaces in toric four-folds. These 16 manifolds are the only ones among the more than half a billion manifolds in the Kreuzer-Skarke list with a non-trivial first fundamental group. We classify the line bundle models on these manifolds, both for $SU(5)$ and $SO(10)$ GUTs, which lead to consistent supersymmetric string vacua and have three chiral families. A total of about 29000 models is found, most of them corresponding to $SO(10)$ GUTs. These models constitute a starting point for detailed heterotic model building on Calabi-Yau manifolds in the Kreuzer-Skarke list. The data for these models can be downloaded \href{http://www-thphys.physics.ox.ac.uk/projects/CalabiYau/toricdata/index.html}{\tt here}.
\end{abstract}
\end{centering}
\newpage

\tableofcontents
\setcounter{footnote}{0}

\section{Introduction}
Over the past few years, a programme of {\em algorithmic string compactification} has been established where a combination of the latest developments in computer algebra and algebraic geometry have been utilized to study the compactification of the heterotic string on smooth Calabi-Yau three-folds with holomorphic vector bundles satsifying the Hermitian Yang-Mills equations \cite{Anderson:2007nc,Anderson:2013xka}.
This is very much in the spirit of the recent advances in applications of algorithmic geometry to string and particle phenomenology \cite{he2012computational,Blumenhagen:2011sq,Blumenhagen:2010ed,Gao:2013pra}. Earlier model building programmes which have paved the way for the current systematic approach have led to a relatively small number of models \cite{Donagi:2004su,Braun:2005nv,Bouchard:2005ag,Anderson:2009mh} which have the particle content of the minimally supersymmetric standard model (MSSM).
In contrast, in the latest scan \cite{Anderson:2013xka} over $10^{40}$ candidate models on complete intersection Calabi-Yau manifolds (CICYs), around $10^{5}$ heterotic standard models were produced.

Of the databases of Calabi-Yau three-folds created over the last three decades in attempting to answer the original question of \cite{Candelas:1987kf} whether superstring theory can indeed give the real world of particle physics, the increasingly numerous - and also chronological - sets are the complete intersections (CICY) in products of projective spaces \cite{Candelas:1987kf}, the elliptically fibred \cite{Friedman:1997yq} and the hypersurfaces in toric four-folds \cite{Kreuzer:2000xy} (cf.~\cite{He:2013epn} for a recent review).
Such Calabi-Yau datasets provide a vast number of candidate internal three-folds for a realistic model, although many of them may be ruled out even on the grounds of basic phenomenology.

The most impressive list, of course, is the last, due to Kreuzer-Skarke (KS).
These total 473,800,776 ambient toric four-folds, each coming from a reflexive polytope in 4-dimensions.
Thus there are at least this many Calabi-Yau three-folds. However, since the majority of the toric ambient spaces are singular and need to be resolved the expected number of Calabi-Yau three-folds from this set is even higher. The Hodge numbers are invariant under this resolution and thus have been extracted to produce the famous plot (which we will exhibit later in the text) of a total of 30,108 distinct Hodge number pairs. To establish stable vector bundles over this largest known set of Calabi-Yau three-folds is of obvious importance.
To truly probe the ``heterotic landscape'' of compactifications which give rise to universes with particle physics akin to ours, one must systematically go beyond the set thus far probed, which had been focused on the CICYs \cite{Anderson:2007nc,Anderson:2013xka,Anderson:2009mh,Anderson:2008uw,Anderson:2009ge,Anderson:2011ns,Anderson:2012yf,Anderson:2011vy,Anderson:2013qca} and the elliptic \cite{Gabella:2008id,Donagi:2004ia} sets.

The study of bundles for model building on the KS dataset was initiated in \cite{He:2009wi} where the Calabi-Yau manifolds with smooth ambient toric four-folds were isolated and studied in detail.
Interestingly, of the some half-billion manifolds, only 124 have smooth ambient spaces.
Bundles which give 3 net generations upon quotienting some potential discrete symmetry and which satisfy all constraints including, notably, Green-Schwarz anomaly cancellation, were classified.

Subsequently, a bench-mark study was performed by going up in $h^{1,1}$ of the KS list \cite{He:2011rs}.
Now, the largest Hodge pairs of any smooth Calabi-Yau three-fold is $(h^{1,1},h^{2,1}) = (491,11)$ (with the mirror having $(h^{1,1},h^{2,1}) = (11,491)$), giving the experimental bound of $960$ on the absolute value of the Euler number.
In \cite{He:2011rs}, we studied the manifolds up to $h^{1,1} = 3$, which already has some 300 manifolds.
The space of positive bundles of monad type were constructed on these spaces.

In any event, the procedure of heterotic compactification is well understood.
Given a generically simply connected Calabi-Yau three-fold $\widetilde X$, we need to find a freely-acting discrete symmetry group $\Gamma$, so that $\widetilde X/\Gamma$ is a smooth quotient. We then need to construct stable $\Gamma$-equivariant bundles $\widetilde V$ on the cover $\widetilde X$ so that on the quotient ${X} =\widetilde X/\Gamma$, $\widetilde V$ descends to a bona fide bundle ${V}$.
It is the cohomology of $V$, coupled with Wilson lines valued in the group $\Gamma$, that gives us the particle content which we need to compute.
In other words, we need to find Calabi-Yau manifolds ${X}$ with non-trivial fundamental group $\pi_1({X}) \simeq \Gamma$.
Often, the manifolds $\widetilde X$ and ${X}$ are referred to as ``upstairs'' and the ``downstairs'' manifolds, to emphasize their quotienting relation.

The simplest set of vector bundles to construct and analyze is that of line bundle sums \cite{Anderson:2011vy,Anderson:2013xka}. Hence, an important step is to classify heterotic line bundle models on Calabi-Yau manifolds in the KS list and extract the ones capable of leading to realistic particle physics. Of course, the existence of freely acting groups $\Gamma$ on the Calabi-Yau manifolds is crucial in order to complete this programme. Unfortunately, these freely-acting symmetries are not systematically known for the KS manifolds.
Indeed, even for the CICY dataset, which had been in existence since the early 1990s, the symmetry groups were only recently classified using the latest computer algebra \cite{Braun:2010vc}. Are there any manifolds in the KS list with known discrete symmetries? A related but simpler question is the following: Are there any manifolds in the KS list already possessing a non-trivial fundamental group?
This latter question was already addressed in Ref.~\cite{Batyrev:2005ts} and the answer is remarkable:
\begin{quote}{\em
Of the some 500 million manifolds in the KS list, only 16 have non-trivial fundamental group.
}\end{quote}

In fact, the 16 covering spaces for these are also in the KS list, and the discrete symmetries $\Gamma$ thereof are known; in particular, their order $|\Gamma|$ is simply the ratio of the Euler numbers of the ``upstairs'' and the ``downstairs'' manifolds.
On these $16$ special ``downstairs'' manifolds one can then directly build stable bundles or, equivalently, stable equivariant bundles can be built on the corresponding $16$ ``upstairs'' manifolds. This is the undertaking of our present paper and constitutes an important scan over a distinguished subset of the KS database.

We emphasize that we expect many more than the aforementioned $16$ manifolds in the KS list to have freely acting symmetries. However, the quotients of those manifolds do not have a description as a hypersurface in a toric four-fold and can, therefore, not be found by searching for non-trivial first fundamental groups in the KS list. Systematic heterotic model building on this full set of KS manifolds with freely-acting symmetries is the challenging task ahead but this will have to await a full classification of freely-acting symmetries. 

The paper is organized as follows.
We start in Section~\ref{sec2} by describing the 16 special base three-folds in detail. In Section~\ref{sec3},
we consider heterotic line bundle models subject to some phenomenological constraints on these manifolds and the algorithm for a systematic scan over all such models is laid out. The result of this scan follows in Section~\ref{sec4} and we conclude with discussion and prospects in Section~\ref{sec5}.

\paragraph*{\bf {Nomenclature}}
Unless stated otherwise, we adhere to the following notations in this paper:
{\renewcommand{\arraystretch}{1.1}
\begin{eqnarray}
&N &\quad \text{The 4-dimensional lattice space of $\Delta$ \nonumber} \\[-.2in]\cr
&M &\quad \text{The dual lattice space of $\Delta^\circ$ \nonumber } \\[-.2in]\cr
&\Delta&\quad \text{Polytope in an auxiliary four-dimensional lattice} \nonumber \\[-.2in]\cr
&\Delta^\circ&\quad\text{Dual polytope of $\Delta$} \nonumber \\[-.2in]\cr 
&\mathcal A_\Delta&\quad\text{``Downstairs'' ambient toric variety constructed from the polytope $\Delta$}\nonumber \\[-.2in]\cr 
&X_\Delta&\quad\text{Calabi-Yau hypersurface three-fold naturally embedded in $\mathcal A_\Delta$} \nonumber \\[-.2in]\cr 
&{\rm Pic}(M)&\quad\text{Picard group of holomorphic line bundles on a manifold $M$}\nonumber \\[-.2in]\cr 
&n&\quad\text{Number of vertices in the polytope $\Delta$} \nonumber \\[-.2in]\cr 
&x_{\rho=1, \cdots, n}&\quad\text{Homogeneous coordinates of an ambient toric variety $\mathcal A$}\nonumber \\[-.2in]\cr 
&D_{\rho=1, \cdots, n}&\quad\text{Divisors defined as the vanishing loci of $x_\rho$}\nonumber \\[-.2in]\cr 
&k&\quad\text{Dimension of Picard group} \nonumber \\[-.2in]\cr 
&J_{r=1, \cdots, k}&\quad\text{Harmonic (1,1)-form basis elements of $H^{1,1}(X, \mathbb Z)$} \nonumber \\[-.2in]\cr 
&\widetilde{\mathcal A_\Delta}&\quad\text{``Upstairs'' ambient toric variety associated with $\mathcal A_\Delta$}\nonumber \\[-.2in]\cr 
&\widetilde{X_\Delta}&\quad\text{Calabi-Yau hypersurface three-fold naturally embedded in $\widetilde{\mathcal A_\Delta}$}\nonumber \\[-.2in]\cr 
&{\rm ch}(V)&\quad\text{Chern character of bundle $V$} \nonumber \\[-.2in]\cr 
&c(V)&\quad\text{Chern class of bundle $V$} \nonumber \\[-.2in]\cr 
&\mu(V)&\quad\text{Mu-slope of bundle $V$} \nonumber \\[-.2in]\cr
&{\rm ind}(V)&\quad\text{Index of the Dirac operator twisted by bundle $V$} \nonumber \\[-.2in]\cr 
&K&\quad\text{K\"ahler cone matrix of a projective variety} \nonumber
\end{eqnarray}
}

\section{The base manifolds: sixteen Calabi-Yau three-folds}\label{sec2}

As mentioned above, the largest known class to date of smooth, compact Calabi-Yau three-folds is constructed as hypersurfaces in a toric ambient four-fold and is often called Kreuzer-Skarke (KS) data set~\cite{Kreuzer:2000xy, Kreuzer:2006um}. 
The huge database consists of the toric ambient varieties $\mathcal A_\Delta$ as well as the Calabi-Yau hypersurfaces $X_\Delta$ therein, both of which are combinatorially described by a ``reflexive'' polytope $\Delta$ living in an auxiliary four-dimensional lattice. 
The classification of reflexive four-polytopes had been undertaken and resulted in the data set of $473,800,766$ polytopes, each of which gives rise to one or more Calabi-Yau three-fold geometries. 

Only $16$ spaces in KS data set carry non-trivial first fundamental groups, which are all of the cyclic form, $\pi_1 \cong \mathbb{Z}/{p\, \mathbb{Z}}$, for $p=2,3,5$~\cite{Batyrev:2005ts}. For the heterotic model-building purposes, one is in need of Wilson lines, so these $16$ Calabi-Yau three-folds form a natural starting point.

More common in heterotic model building is to start from a simply-connected Calabi-Yau three-fold $\widetilde X$ with freely-acting discrete symmetry group $\Gamma$ and then form the quotient $X=\widetilde X/\Gamma$ which represents a Calabi-Yau manifold with first fundamental group equal to $\Gamma$. Indeed, for the CICY data set~\cite{Candelas:1987kf}, all the 7890 Calabi-Yau three-folds turn out to be simply-connected and a heavy computer search had to be performed to classify the freely-acting discrete symmetries~\cite{Braun:2010vc}. Typical heterotic models have thus been built firstly on the upstairs CICY $\widetilde X$ and have then been descended to the downstairs Calabi-Yau $X$. A similar approach has also been taken for the model building based on the KS list carried out in Ref.~\cite{He:2011rs}. 

In this paper, we attempt to construct heterotic models outright from the downstairs geometry.
We shall start in this section by describing some basic geometry of the sixteen toric Calabi-Yau three-folds $X$ with $\pi_1(X)\neq \emptyset$. 
This includes Hodge numbers, Chern classes, intersection rings and K\"ahler cones.  
The precise quotient relationship with the corresponding upstairs three-folds $\widetilde X$, as well as the full list of relevant geometries, can be found in Appendix~\ref{disofud}.

\subsection{The construction}\label{2.1}

Let us label the sixteen Calabi-Yau three-folds and their ambient toric four-folds by $X_{i=1, \cdots, 16}$ and $\mathcal A_{i=1, \cdots, 16}$, respectively. 
They come from the corresponding (reflexive) polytopes $\Delta_i$ in an auxiliary rank-four lattice $N$, whose vertex information~\cite{Batyrev:2005ts} is summarised in Appendix~\ref{down3}. 
Before describing their geometry in section~\ref{sec22}, partly to set the scene up, we illustrate the general procedure for the toric construction of Calabi-Yau three-fold, by the explicit example, $X_3 \subset \mathcal A_3$ and $\Delta_3$. For a more detailed introduction, interested readers are kindly referred, e.g., to Ref.~\cite{He:2009wi} and references therein. 

Let us first extract the lattice polytope $\Delta_3$ from Appendix~\ref{down3}: 
$$\begin{array}{l}
\left( {\begin{array}{*{32}{c}}
{{x_1}}&{{x_2}}&{{x_3}}&{{x_4}}&{{x_5}}&{{x_6}}&{{x_7}}&{{x_8}} \\ \hline
2&{ 0}&{0}&{ 0}&{ 0}&{0}&{0}&{-2}\\
{0}&-1&0&1&-1&0&1&0\\
{ 0}&0&-1&1&-1&1&0&0\\
{1}&0&0&1&-1&0&0&-1
\end{array}} \right) \ .
\label{example3}
\end{array}$$
It has $n=8$ vertices in $N \simeq \mathbb Z^4$ leading to $8$ homogeneous coordinates $x_{\rho=1, \cdots, 8}$ for the ambient toric four-fold $\mathcal A_3$; 
the $4$ rows of the above matrix describe the $4$ projectivisations that reduce the complex dimension from $8$ down to $4$. 
Next, the dual polytope $\Delta_3^\circ$ in the dual lattice $M$ is constructed as 
$$\Delta_3^\circ:=\{ m \in M~|~\langle m,v\rangle  \geqslant  - 1\ , ~\forall v \in \Delta_3\} \ , $$
and one can easily check that $\Delta_3^\circ$ is also a lattice polytope. 
Then it so turns out that each of the lattice points in $\Delta_3^\circ$ is mapped to a global section of the normal bundle for the the embedding, $X_3 \subset \mathcal A_3$, of the Calabi-Yau three-fold (see Eq.~(45) of Ref.~\cite{He:2009wi} for the explicit map). 
Here, $\Delta_3^\circ$ has $41$ lattice points and the corresponding $41$ sections are obtained as: 
{\small\begin{equation}
\begin{gathered}
~~x_2^2x_3^2x_4^2x_8^2~,~~x_2^2x_3^2x_5^2x_8^2~,~~{x_1}x_2^2x_3^2{x_4}{x_5}{x_8}~,~~x_1^2x_2^2x_3^2x_4^2~,~~x_2^2{x_3}{x_4}{x_5}{x_6}x_8^2~,~~{x_1}x_2^2{x_3}x_4^2{x_6}{x_8}~, \hfill \\
~~x_2^2x_4^2x_6^2x_8^2~,~~{x_2}x_3^2{x_4}{x_5}{x_7}x_8^2~,~~{x_1}{x_2}x_3^2x_4^2{x_7}{x_8}~,~~{x_2}{x_3}x_4^2{x_6}{x_7}x_8^2~,~~x_3^2x_4^2x_7^2x_8^2~,~~x_1^2x_2^2x_3^2x_5^2~, \hfill \\
{x_1}x_2^2{x_3}x_5^2{x_6}{x_8}~,~~x_1^2x_2^2{x_3}{x_4}{x_5}{x_6}~,~~x_2^2x_5^2x_6^2x_8^2~,~~{x_1}x_2^2{x_4}{x_5}x_6^2{x_8}~,~~x_1^2x_2^2x_4^2x_6^2~,~~{x_1}{x_2}x_3^2x_5^2{x_7}{x_8}~, \hfill \\
~~x_1^2{x_2}x_3^2{x_4}{x_5}{x_7}~,~~{x_2}{x_3}x_5^2{x_6}{x_7}x_8^2~,~~{x_1}{x_2}{x_3}{x_4}{x_5}{x_6}{x_7}{x_8}~,~~ x_1^2{x_2}{x_3}x_4^2{x_6}{x_7}~,~~{x_2}{x_4}{x_5}x_6^2{x_7}x_8^2~, \hfill \\
{x_1}{x_2}x_4^2x_6^2{x_7}{x_8}~,~~x_3^2x_5^2x_7^2x_8^2~,~~{x_1}x_3^2{x_4}{x_5}x_7^2{x_8}~,~~x_1^2x_3^2x_4^2x_7^2~,~~{x_3}{x_4}{x_5}{x_6}x_7^2x_8^2~,~~{x_1}{x_3}x_4^2{x_6}x_7^2{x_8}~, \hfill \\
~~x_4^2x_6^2x_7^2x_8^2~,~~x_1^2x_2^2x_5^2x_6^2~,~~x_1^2{x_2}{x_3}x_5^2{x_6}{x_7}~,~~{x_1}{x_2}x_5^2x_6^2{x_7}{x_8}~,~~x_1^2{x_2}{x_4}{x_5}x_6^2{x_7}~,~~ x_1^2x_3^2x_5^2x_7^2~, \hfill \\
~~{x_1}{x_3}x_5^2{x_6}x_7^2{x_8}~,~~x_1^2{x_3}{x_4}{x_5}{x_6}x_7^2~,~~x_5^2x_6^2x_7^2x_8^2~,~~{x_1}{x_4}{x_5}x_6^2x_7^2{x_8}~,~~x_1^2x_4^2x_6^2x_7^2~,~~x_1^2x_5^2x_6^2x_7^2 \ . \hfill \\ 
\end{gathered} 
\end{equation}} 
which, when linearly combined, give the defining equation for $X_3$.

Note that as the non-trivial fundamental group is torically realised, it is natural to expect that the KS list also contains the sixteen upstairs geometries, which we denote by $\widetilde{X_i} \subset \widetilde {\mathcal A_i}$. 
By construction, the upstairs three-folds $\widetilde {X_i}$ should admit a freely-acting discrete symmetry $\Gamma_i$ so that $X_i = \widetilde {X_i} /\Gamma_i$ with $\pi_1(X_i) = \Gamma_i$. 
We have indeed found the corresponding upstairs polytopes $\widetilde{\Delta_i}$ associated with the sixteen downstairs (see Appendix~\ref{down3} for their vertex lists). 
It turns out that three of the sixteen upstairs Calabi-Yau three-folds $\widetilde{X_i} \subset \widetilde{\mathcal A_i}$ belong to the CICY list~\cite{Candelas:1987kf}: $\widetilde{X_1}$ is the quintic three-fold in $\mathbb{P}^4$, $\widetilde{X_2}$ the bi-cubic in $\mathbb{P}^2\times\mathbb{P}^2$ and $\widetilde {X_3}$ the tetra-quadric in ${\mathbb{P}^1}^{\times 4}$. 
Although the models in this paper are constructed over the downstairs manifolds, one can compare, as a cross-check, the models over $X_1, X_2$ and $X_3$ with the known results over the CICYs~\cite{Anderson:2011ns,Anderson:2012yf}.

We finally remark that the ambient toric varieties $\mathcal A_\Delta$ constructed by the standard toric procedure might in general involve singularities. 
In order to obtain smooth Calabi-Yau hypersurfaces $X$, one must resolve the singularities of the ambient space to a point-like level via ``triangulation'' of the polytope $\Delta$ in a certain manner~\cite{cox2011toric}. 
The triangulation splits $\Delta$ maximally and leads to a partial desingularisation of the toric variety $\mathcal A_\Delta$.  
In principle, there may arise several different desingularisations for a single toric variety $\mathcal A_\Delta$, in which case the number of geometries increases. 
Indeed, $X_6$ and $X_{14}$ turn out to have two and three desingularisations, respectively, while the other fourteen Calabi-Yau manifolds only have one each.

\subsection{Some geometrical properties} \label{sec22}
Having constructed the Calabi-Yau three-folds in the previous subsection, we now move on to study their geometrical properties relevant to the heterotic model-building. Instead of describing all the details in an abstract manner, we continue with the example $X_3$; the $\mathbb Z_2$-quotient of the tetra-quadric $\widetilde X_3$ in ${\mathbb{P}^1}^{\times 4}$. The detailed prescription for computing the geometric properties can be found from Appendix~B~of~\cite{He:2009wi}. Alternatively, one could also make use of the computer package PALP~\cite{PALP} to extract all the information. The resulting geometry can be summarised as follows.

Firstly, we have $k\equiv{\rm rk}({\rm Pic}(\mathcal A_3))= 4$ and hence, the Picard group is generated by four elements $J_{r=1, \cdots, 4}$. 
One can then choose the basis elements appropriately so that the toric divisors $D_{\rho=1, \cdots, 8}$ defined as the vanishing locus of the homogeneous coordinate $x_\rho$ have the following expressions:
\begin{equation} \label{down3int}
D_1=J_4, ~D_2=J_3, ~D_3=J_2, ~D_4=J_1, ~D_5=J_1, ~D_6=J_2, ~D_7=J_3, ~D_8=J_4 \ ,
\end{equation}
where, by abuse of notation, the harmonic $(1,1)$-forms $J_r$ are also used to denote the basis of Picard group.
Furthermore, unless ambiguities arise, we shall not attempt to carefully distinguish the harmonic forms of the ambient space from their pullbacks to the hypersurface.
Next, the intersection polynomial of $X_3$ is: $$J_1\, J_2\, J_3+J_1\, J_2\, J_4+J_1\, J_3\, J_4+J_2\, J_3\, J_4 \ , $$ which means that the only non-vanishing triple intersections are $$d_{123}({X_3})=d_{124}({X_3})=d_{134}({X_3})=d_{234}({X_3})=1 \, $$ 
and those obtained by the permutations of the indices above. 
The Hodge numbers can also be easily computed:
$$\begin{array}{*{20}{c}}
{{h^{1,1}({X_3})} = 4,}&{{h^{1,2}({X_3})} = 36} \ ,
\end{array}$$
leading to the Euler character $\chi({X_3})= -64$.
The second Chern character for the tangent bundle, which is crucial for the anomaly check, is given by
\begin{equation}
{\rm ch}_2(TX)=\{12,12,12,12\} = \sum\limits_{r=1}^4 12 \;\nu^r \ , 
\end{equation}
in the dual $4$-form basis $\nu^{r=1, \cdots, 4}$ defined such that $\int_{X_3} J_r \wedge \nu^s = \delta_r^s$.
Finally, the K\"ahler cone matrix $K= \left[K_{rs}\right]$, describing the K\"ahler cone as the set of all K\"ahler parameters $t^r$ satysfying $K_{rs}t^s\geq 0$ for all $r=1,\ldots ,h^{1,1}(X)$, takes the form
\begin{equation}
K=\left(
\begin{array}{cccc}
 1 & 0 & 0 & 0 \\
 0 & 1 & 0 & 0 \\
 0 & 0 & 1 & 0 \\
  0 & 0 & 0 & 1 \\
\end{array}
\right) \; ,
\end{equation}
thus representing the part of $\bf t$ space with $t^{r=1,\cdots, 4}>0$.

The reader might have notice that $h^{1,1}(X_3)=4 = h^{1,1}(\mathcal A_3)$ in this example.
In general, however, $h^{1,1}(X)$ can be larger than $h^{1,1}(\mathcal A)$ and a hypersurface of this type is called ``non-favourable,'' as we do not have a complete control over all the K\"ahler forms of $X$ through the simple toric description of the ambient space $\mathcal A$. 
The notion of favourability means that the K\"ahler structure of the Calabi-Yau hypersurface is entirely descended down from that of the ambient space; namely, the integral cohomology group of the hypersurface can be realised by a toric morphism from the ambient space.
Amongst the sixteen downstairs geometries $X_i$, only the two, $X_{15}$ and $X_{16}$, turn out to be non-favourable.  
As we do not completely understand their K\"ahler structure, we will not attempt to build models on either of these two manifolds.

In Appendix~\ref{sumBG}, the geometrical properties summarised so far for $X_3 \subset \mathcal A_3$ are tabulated for all the downstairs manifolds $X_i \subset \mathcal A_i$, as well as their upstairs covers $\widetilde{X_i} \subset \widetilde{\mathcal A_i}$, $i=1, \dots, 16$. Another illustration for how to read off the geometry from the table is given in Appendix~\ref{monoexample} for $X_1 \subset \mathcal A_1$ and $\widetilde{X_1} \subset \widetilde{\mathcal A_1}$.

Let us close this subsection by touching upon an issue with multiple triangulations. 
As mentioned in section~\ref{2.1}, the Calabi-Yau three-folds $X_6$ and $X_{14}$ turn out to admit two and three triangulations, respectively. Here we take the former as an example. 
Its toric data is encoded in the polytope $\Delta_6$:
$$\begin{array}{l}
\left( {\begin{array}{*{28}{c}}
{{x_1}}&{{x_2}}&{{x_3}}&{{x_4}}&{{x_5}}&{{x_6}}&{{x_7}} \\ \hline
-4&0&0& 0&2&0&-2\\
-3&1&0&-1&0&-2&-2\\
1&0&1&-1&0&-1&0\\
-1&0&0&-1&1&0&-1
\end{array}} \right) \ ; 
\end{array}$$
this polytope turns out to admit the following two different star triangulations \footnote{A triangulation is star if all maximal simplices contain a common point, in this case reduced to be cones expanded by four vertices and the origin point. In our notation the origin point is omitted, leaving only the four indices labeling the vertices.},
\begin{eqnarray} \nonumber 
\mathcal T_1 &=&  \small 
\begin{gathered}
  \{ \{ 1, 2, 5, 6\} , \{ 2, 3, 4, 5\} , \{ 1, 2, 3, 5\} , \{ 2, 4, 5, 6\} , \{ 2, 4, 6, 7\} , \{ 1, 2, 6, 7\} , \hfill \\
  \{ 2, 3, 4, 7\} , \{ 1, 2, 3, 7\} , \{ 3, 4, 6, 7\} , \{ 1, 3, 6, 7\} , \{ 3, 4, 5, 6\} , \{ 1, 3, 5, 6\} \}  \hfill \\ 
\end{gathered} \  \\  \nonumber
\mathcal T_2&=& \small 
\begin{gathered}
  \{ \{ 1, 2, 5, 6\} , \{ 2, 3, 4, 5\} , \{ 1, 2, 3, 5\} , \{ 2, 4, 5, 6\} , \{ 2, 4, 6, 7\} , \{ 1, 2, 6, 7\}, \hfill \\
  \{ 2, 3, 4, 7\} , \{ 1, 2, 3, 7\} , \{ 4, 5, 6, 7\} , \{ 1, 5, 6, 7\} , \{ 3, 4, 5, 7\} , \{ 1, 3, 5, 7\} \} \hfill \\ 
\end{gathered} \
\end{eqnarray}
where triangulations of the polytope $\Delta_6$ are described as a list of four-dimensional cones. For instance, the first element $\{1,2,5,6\} \in \mathcal T_1$ represents the four-dimensional cone spanned by the corresponding four vertices: 
$$(-4,-3,1,-1),~(0,1,0,0),~(2,0,0,1),~(0,-2,-1,0).$$ 
It also turns out that the two smooth hypersurfaces, associated with the two triangulations $\mathcal T_1$ and $\mathcal T_2$, have the same intersection structure and the same second Chern class. 
It is expected in such a case that the two Calabi-Yau hypersurfaces are connected in the K\"ahler moduli space.
In other words, the two K\"ahler cones adjoin along a common facet.  
Thus, the pair can be thought of as leading to a single Calabi-Yau three-fold $X_6$, whose K\"ahler cone is the union of the two sub-cones,
\[K({X_6}) = \bigcup\limits_{j=1}^2 {{K_j}} \ , \]  
where $K_1$ and $K_2$ are the K\"ahler cones of the two hypersurfaces associated with $\mathcal T_1$ and $\mathcal T_2$, respectively (see Ref.~\cite{He:2011rs} for the details). 
The K\"ahler cone matrices for the two sub-cones turn out to be
$$
{K_1} = \left( {\begin{array}{*{20}{r}}
  0&1&0 \\ 
  1&0&{ - 2} \\ 
  0&{ - 1}&1 
\end{array}} \right) ~~\text{and~~} {K_2} = \left( {\begin{array}{*{20}{r}}
  0&0&1 \\ 
  1&0&{ - 2} \\ 
  0&1&{ - 1} 
\end{array}} \right) \; , 
$$
and therefore, the K\"ahler cone matrix for the union can be computed as:
$$
K({X_6}) = \left( {\begin{array}{*{20}{c}}
  1&0&0 \\ 
  0&1&0 \\ 
  0&0&1 
\end{array}} \right) \ . 
$$

One can similarly play with $\Delta_{14}$. For this geometry as well it turns out that the three triangulations lead to a single Calabi-Yau three-fold, $X_{14}$. 
As for the K\"ahler cone, the three sub-cones are
\begin{equation}
{K_1} = \left( {\begin{array}{*{20}{r}}
  0&1&0 \\ 
  1&-1&{0} \\ 
  0&{ - 1}&1 
\end{array}} \right) ~~\text{~~} {K_2} = \left( {\begin{array}{*{20}{r}}
  0&0&1 \\ 
  0&1&{ - 1} \\ 
  1&0&{ - 1} 
\end{array}} \right) ~~\text{~~} {K_3} = \left( {\begin{array}{*{20}{r}}
  1&0&0 \\ 
  -1&0&{ - 1} \\ 
  -1&1&{ 0} 
\end{array}} \right) \; ,
\end{equation} 
and via the simple joining one obtains the K\"ahler cone of $X_{14}$:
\begin{equation}
K({X_{14}}) = \left( {\begin{array}{*{20}{c}}
  1&0&0 \\ 
  0&1&0 \\ 
  0&0&1 
\end{array}} \right) \ . 
\end{equation}
In summary, although there are different triangulations for $\Delta_6$ and $\Delta_{14}$, one ends up obtaining a single geometry each, $X_6$ and $X_{14}$, respectively.

\subsection{Location in the Calabi-Yau landscape}
\begin {figure}[H]
\centering
\includegraphics[scale=0.38]{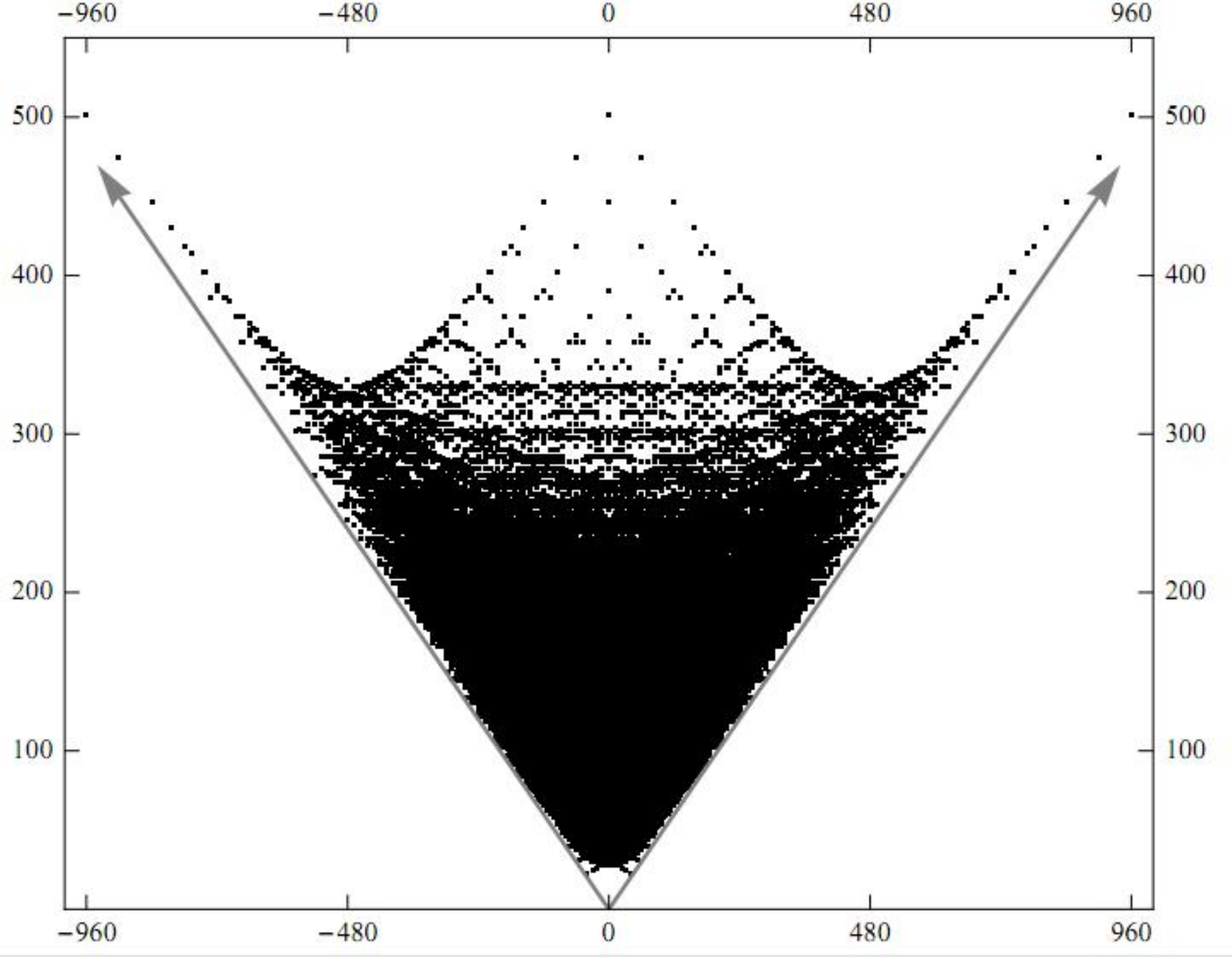}~~~
\includegraphics[scale=0.24]{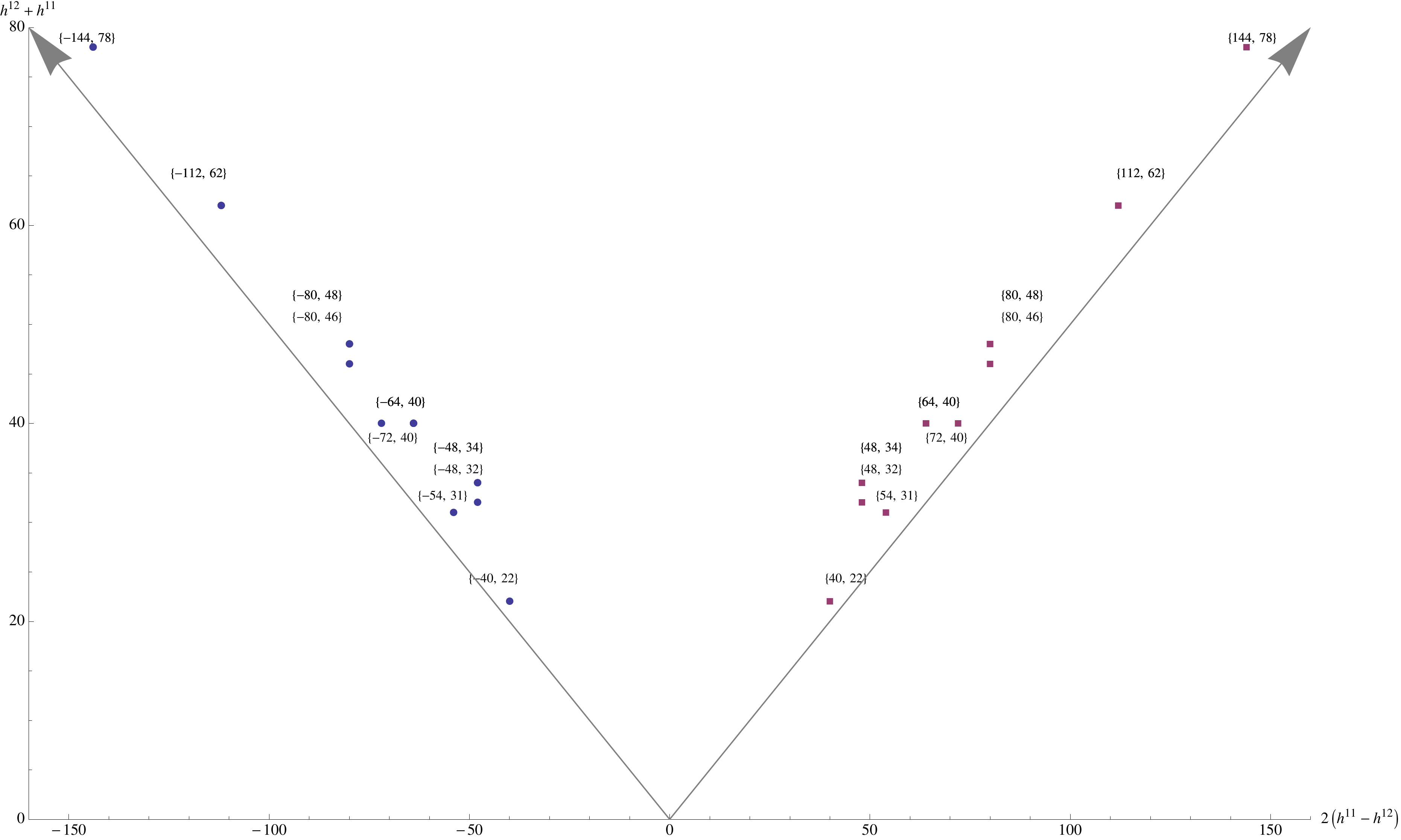}
\caption{\em The Hodge number plot: $\{2(h^{1,1}-h^{2,1}),h^{1,1}+h^{2,1}\}$. The left figure is for all the Calabi-Yau three-folds known to date and the right is for the sixteen non-simply-connected Calabi-Yau three-folds $X_i$ as well as their mirrors; the blue round dots are for the original sixteen and the purple squares are for the mirrors.}\label{figeula}
\end{figure}

Since a very special corner in the landscape of Calabi-Yau three-folds has been chosen, it might be interesting to see the location of these sixteen, say, in the famous Hodge number plot~\cite{candelas2012abundance}. 
Figure~\ref{figeula} shows the Hodge number plot of all the Calabi-Yau three-folds known to date, together with that of the sixteen manifolds $X_i$ and of their mirrors.
Some basic topological data for both downstairs $X_i$ and upstairs $\widetilde {X_i}$ is also summarized in Table~\ref{t:the16} for reference.

\begin{table}[h!t!]
{\renewcommand{\arraystretch}{1.25}
\begin{footnotesize}
\resizebox{\textwidth}{!}{$
\begin{array}{|c||c|c|c|c|c|c|c|c|c|c|c|c|c|c|c|c|}
\hline 
i &~~1~~&~~2~~&~~3~~&~~4~~&~~5~~&~~6~~&~~7~~&~~8~~&~~9~~&~~10~~&~~11~~&~~12~~&~~13~~&~~14~~&~~15~~&~~16~~\\ \hline \hline
h^{1,1}(\widetilde{X_i}) &1 &2& 4& 4_{\rm nf}& 3& 3& 4_{\rm nf}& 4_{\rm nf} &4 &4& 4& 5_{\rm nf} &5_{\rm nf} &3 &7_{\rm nf}& 7_{\rm nf}
\\ \hline
h^{1,1}(X_i)&1 &2 &4 &2 &3 &3 &3 &3 &4 &4 &4 &4 &4 &3 &5_{\rm nf}& 5_{\rm nf}
\\ \hline \hline
-\chi(\widetilde{X_i}) & 200 &162 &128 &216& 160& 224& 288& 288 &96 &128& 128& 160& 160& 224& 96& 96
\\ \hline
-\chi(X_i) &40 &54 &64& 72& 80& 112& 144& 144& 48& 64& 64& 80& 80& 112& 48 &48
\\ \hline \hline
\pi_1(X_i)& \mathbb Z_5 & \mathbb Z_3 & \mathbb Z_2 &\mathbb Z_3 & \mathbb Z_2& \mathbb Z_2& \mathbb Z_2& \mathbb Z_2& \mathbb Z_2& \mathbb Z_2& \mathbb Z_2& \mathbb Z_2& \mathbb Z_2& \mathbb Z_2& \mathbb Z_2& \mathbb Z_2
\\ \hline
\end{array}
$}
\end{footnotesize}
\begin{center}
\caption{\em Picard numbers and Euler characters of the downstairs Calabi-Yau three-folds $X_i$ and their upstairs covers $\widetilde{X_i}$, for $i=1, \dots, 16$. In the last row is also shown the $\pi_1$ of the downstairs manifolds $X_i$. The subscript ``nf'' for Picard number indicates that the geometry is non-favourable.}
\label{t:the16}
\end{center}
}
\end{table}

\section{Physical constraints and search algorithm }\label{sec3}

As indicated in Table~\ref{t:the16}, of the sixteen downstairs three-folds, the first fourteen, $X_{i=1, \cdots, 14}$, turn out to be favourable and, in this paper, we shall take the initial step towards the construction of heterotic line bundle standard models on them. The main difficulty with the two non-favourable geometries 
arises from the K\"ahler forms which do not descend from the ambient space; the corresponding components of the K\"ahler matrix and the triple intersection numbers are difficult to obtain from the ambient toric data, since the line bundles could be safely descended down to CY manifolds are coming only from toric divisors, with a smaller number than the dimension of CY manifold. While for the extra line bundles of CY manifolds, it is not straight forward to write them out and not possible to compute the triple intersection numbers since the calculation is essentially done over the toric variety. Therefore, the missing info makes it impossible to fully check certain consistency conditions of the bundle, notably the poly-stability condition discussed below.

\subsection{Choice of bundles and gauge group}

Let us begin by discussing the choice of gauge bundle and the resulting four-dimensional gauge group. First of all, we need to choose a bundle $V$ with structure group $G$ which embeds into the visible $E_8$ gauge group. The resulting low-energy gauge group, $H$, is the commutant of $G$ within $E_8$. As discussed earlier, for $V$ we would like to consider Whitney sums of line bundles of the form
\begin{equation} 
 V=\bigoplus_{a=1}^nL_a\; ,\quad L_a={\cal O}_X({\bf k}_a)\; , \label{lbs}
\end{equation} 
where the line bundles are labeled by integer vectors ${\bf k}_a$ with $h^{1,1}(X)$ components $k_a^r$ such that their first Chern classes can be written as $c_1(L_a)=k_a^rJ_r$. The structure group of this line bundle sum should have an embedding into $E_8$. For this reason, we will demand that $c_1(V)=0$ or, equivalently,
\begin{equation}
 \sum_{a=1}^n{\bf k}_a=0\; , \label{c10}
\end{equation}
which leads, generically, to the structure group $G=S(U(1)^n)$. For $n=4,5$ this structure group embeds into $E_8$ via the subgroup chains $S(U(1)^4)\subset SU(4)\subset E_8$ and $S(U(1)^5)\subset SU(5)\subset E_8$, respectively. This results in the commutants $H=SO(10)\times U(1)^3$ for $n=4$ and $H=SU(5)\times U(1)^4$ for $n=5$. Both, $SU(5)$ and $SO(10)$, are attractive grand unification groups and they can be further broken to the standard model group after the inclusion of Wilson lines. Hence, constructing such $SU(5)$ and $SO(10)$ models, subject to further constraints discussed below, is the first step in the standard heterotic model building programme. The additional $U(1)$ symmetries turn out to be typically Green-Schwarz anomalous. Hence, the associated gauge bosons are super massive and of no phenomenological concern.

\subsection{Anomaly cancelation}\label{gencons}

In general, anomaly cancelation can be expressed as the topological condition
\begin{equation}\label{anomaly1}
 {\rm ch}_2(V) + {\rm ch}_2(\hat V)-{\rm ch}_2(TX)  = [C] \; , 
\end{equation}
where $V$ is the bundle in the observable $E_8$ sector, as discussed, $\tilde{V}$ is its hidden counterpart and $[C]$ is the homology class of a holomorphic curve, $C$, wrapped by a five-brane. A simple way to guarantee that this condition can be satisfied is to require that 
\begin{equation}
{\rm c}_2(TX) - {\rm c}_2(V) \in {\rm Mori}(X) \ , \label{an2}
\end{equation}
where ${\rm Mori}(X)$ is the cone of effective classes of $X$. Here, we have used that ${\rm ch}_2(TX)=-{\rm c}_2(TX)$ and that ${\rm ch}_2(V)=-c_2(V)$ for bundles $V$ with $c_1(V)=0$. Provided condition~\eqref{an2} holds the model can indeed always be completed in an anomaly-free way so that Eq.~\eqref{anomaly1} is satisfied. Concretely, Eq.~\eqref{an2} guarantees that there exists a complex curve $C$ with $[C]={\rm c}_2(TX) - {\rm c}_2(V)$, so that wrapping a five brane on this curve and choosing the hidden bundle to be trivial will do the job (although other choices involving a non-trivial hidden bundle are usually possible as well). 

To compute the the second Chern class $c_2(V)=c_{2r}(V)\nu^r$ of line bundle sums~\eqref{lbs} we can use the result
\begin{equation}
c_{2r}(V)=-\frac{1}{2}d_{rst}\sum_{a=1}^nk_a^sk_a^t\; , \label{c2V}
\end{equation}
where $d_{rst}$ are the triple intersection numbers. For the $16$ manifolds under consideration these numbers, as well as the second Chern classes, $c_2(TX)$, of the tangent bundle are provided in Appendix~\ref{disofud}.
\subsection{Poly-stability}\label{polystab}
The Donaldson-Uhlenbeck-Yau theorem states that for a ``poly-stable'' holomorphic vector bundle $V$ over a K\"ahler manifold $X$, there exists a unique connection satisfying the Hermitian Yang-Mills equations. Thus, in order to make the models consistent with supersymmetry, we need to verify that the sum of holomorphic line bundles is poly-stable.

Poly-stability of a bundle (coherent sheaf) $\mathcal F$ is defined by means of the {\it slope}
\begin{equation} \label{poly_stable}
\mu (\mathcal F) \equiv \frac{1}{{\rm rk}(\mathcal F)}\int_X {{c_1}(\mathcal F) \wedge J \wedge J} \ ,  
\end{equation}
where $J$ is the K\"ahler form of the Calabi-Yau three-fold $X$. 
The bundle $\mathcal F$ is called {\it poly-stable} if it decomposes as a direct sum of stable pieces, 
\begin{equation}
\mathcal F = \bigoplus\limits_{a=1}^m \mathcal F_a \ , 
\end{equation} 
of equal slope $\mu (\mathcal F_a) = \mu (\mathcal F)$, for $a=1, \cdots, m$. 
In our case, the bundle $V$ splits into the line bundles $L_a$ as in Eq.~\eqref{lbs}. 
Line bundles, however, are trivially stable as they do not have a proper subsheaf. This feature is one of the reasons why heterotic line bundle models are technically much easier to deal with than models with non-Abelian structure groups. All that remains from poly-stability is the conditions on the slopes. Since $c_1(V)=0$, we have $\mu(V)=0$ and, hence, the slopes of all constituent line bundles $L_a$ must vanish. This translates into the conditions
\begin{equation}
\mu(L_a)=k_a^r\kappa_r = 0  \quad \text{where} \quad \kappa_r = d_{rst}t^{s}t^{t}\; , \label{slope0}
\end{equation}
for $a=1,\ldots ,n$ which must be satisfied simultaneously for K\"ahler parameters $t^r$ in the interior of the K\"ahler cone. The intersection numbers and the data describing the K\"ahler cone for our $16$ manifolds is provided in Appendix~\ref{disofud}.

\subsection{$SU(5)$ GUT theory}

A model with a (rank four or five) line bundle sum~\eqref{lbs} in the observable sector that satisfies the constraints~\eqref{c10}, \eqref{an2} and \eqref{slope0} can be completed to a consistent supersymmetric heterotic string compactification leading to a four-dimensional $N=1$ supergravity with gauge group $SU(5)$ or $SO(10)$ (times anomalous $U(1)$ factors). Subsequent conditions, which we will impose shortly, are physical in nature and are intended to single out models with a phenomenologically attractive particle spectrum. The details of how this is done somewhat depend on the grand unified group under consideration and we will discuss the two cases in turn, starting with $SU(5)$. 

In this case we start with a line bundle sum~\eqref{lbs} of rank five ($n=5$) and associated structure group $G=S(U(1)^5)$. This leads to a four-dimensional gauge group $H=SU(5)\times S(U(1)^5)$. 
The four-dimensional spectrum consists of the following $SU(5)\times S(U(1)^5)$ multiplets:
\begin{equation}
 {\bf 10}_a\;,\;\; \overline{\bf 10}_a\;,\;\;\overline{\bf 5}_{a,b}\;,\;\;{\bf 5}_{a,b}\;,\;\;{\bf 1}_{a,b}\; .
\end{equation}
Here, the subscripts $a,b,\dots =1,\ldots ,5$ indicate which of the additional $U(1)$ factors in $S(U(1)^5)$ the multiplet is charged under. A $ {\bf 10}_a$ ($\overline{\bf 10}_a$) multiplet carries charge $1$ ($-1$) under the $a^{\rm th}$ $U(1)$ and is uncharged under the others.
A $\overline{\bf 5}_{a,b}$ (${\bf 5}_{a,b}$), where $a<b$, carries charge $1$ ($-1$) only under the $a^{\rm th}$ and $b^{\rm th}$ $U(1)$ while the only charges of a singlet ${\bf 1}_{a,b}$, where $a\neq b$, are $1$ under the $a^{\rm th}$ $U(1)$ and $-1$ under the $b^{\rm th}$ $U(1)$. 

The multiplicity of these various multiplets is computed by the dimension of associated cohomology groups as given in Table~\ref{t:spectrum}.
\begin{table}[h!t!]
{\renewcommand{\arraystretch}{1.1}
\[
\begin{array}{|c|c|c|}
\hline 
\mbox{~$SU(5)\times S(U(1)^5)$ repr.} & \mbox{~~associated cohomology~~} & \mbox{~~contained in~~}\\
\hline \hline 
\bold{10}_a & H^{1}(X,L_a)& H^1(X,V) \\\hline
\bold{\overline {10}}_a & H^{1}(X,L_a^*) &H^1(X,V^*)\\\hline
\bold{\overline {5}}_{a,b} &H^1(X,L_a\otimes L_b) & H^1(X,\wedge^2 V) \\\hline
\bold{5}_{a,b} &  H^1(X,L^*_a \otimes L^*_b) & H^1(X,\wedge^2 V^{*}) \\\hline
\bold{1}_{a,b} & H^1(X,L_a \otimes L_b^*) &H^1(X,V \otimes V^{*})  \\\hline
\end{array}
\]}
\begin{center}
\caption{\em The spectrum of $SU(5)$ models and associated cohomology groups. }
\label{t:spectrum}
\end{center}
\end{table}
The most basic phenomenological constraint to impose on this spectrum is chiral asymmetry of three in the ${\bf 10}$--$\overline{\bf 10}$ sector. This translates into the condition $${\rm ind}(V)=-3 \ , $$  on the index of $V$ which can be explicitly computed from
\begin{equation}
 {\rm ind}(V)=\frac{1}{6}d_{rst}\sum_{a=1}^nk_a^rk_a^sk_a^t\; . \label{10cond}
\end{equation} 
Of course, a similar constraint on the chiral asymmetry should hold in the $\overline{\bf 5}$--${\bf 5}$ sector. In general, for a rank $m$ bundle $V$, we have the relation 
\begin{equation}\label{v2andv1}
{\rm ind}(\wedge^2 V) = (m-4) {\rm ind}(V)
\end{equation} 
So for the rank five bundles presently considered it follows that ${\rm ind}(\wedge^2 V) = {\rm ind}(V)$. Hence the requirement~\eqref{10cond} on the chiral asymmetry in the ${\bf 10}$--$\overline{\bf 10}$  sector already implies the correct chiral asymmetry for the $\overline{\bf 5}$--${\bf 5}$ multiplets, ${\rm ind}(\wedge^2 V)=-3$, and no additional constraint is required. 

The index constraints imposed so far are necessary but of course not sufficient for a realistic spectrum. For example, one obvious additional phenomenological requirement would be the absence of $\overline{\bf 10}$ multiplets which amounts to the vanishing of the associated cohomology group, that is, $h^1(X,V^*)=0$.  However, cohomology calculations are much more involved than index calculations and currently there is no complete algorithm for calculating line bundle cohomology on Calabi-Yau hypersurfaces in toric four-folds. For this reason, we will not impose cohomology constraints on our models in the present paper, although this will have to be done at a later stage. 

However, working with line bundle sums allows us to impose slightly stronger constraints which are based on the indices of the individual line bundles. Of course we can express the indices of $V$ and $\wedge^2 V$ in terms of the indices of their constituent line bundles as
\begin{equation}
 {\rm ind}(V)=\sum_{a=1}^n{\rm ind}(L_a)\;,\quad {\rm ind}(\wedge^2V)=\sum_{a<b}{\rm ind}(L_a\otimes L_b)\; , \label{Vsum}
\end{equation} 
where, by the index theorem, the index of an individual line bundle $L={\cal O}_X({\bf k})$ is given by
\begin{equation}\label{ind-linebundle}
 {\rm ind}({L})=d_{rst}\left(\frac{1}{6}k^r k^s k^t +\frac{1}{12} k^r c_2^{st}(TX) \right) \; .
\end{equation}
Suppose that ${\rm ind}(L_a)>0$ for one of the line bundles $L_a$. Then, in this sector, there is a chiral net-surplus of $\overline{\bf 10}$ multiplets which is protected by the index and will survive the inclusion of a Wilson line. Since such $\overline{\bf 10}$ multiplets and their standard-model descendants are phenomenologically unwanted we should impose~\footnote{The caveat is that line bundle models frequently represent special loci in a larger moduli space of non-Abelian bundles. Line bundle models with exotic states -- vector-like under the GUT group/standard model group but chiral under the $U(1)$ symmetries -- may become realistic when continued into the non-Abelian part of the moduli space where some or all of the $U(1)$ symmetries are broken. In this case, the exotic states may become fully vector-like, acquire a mass and are removed from the low-energy spectrum. While this is an entirely plausible model building route, here we prefer a ``cleaner'' approach where the spectrum at the Abelian locus can already lead to a realistic spectrum.} that ${\rm ind}(L_a)\leq 0$ for all $a$. Combining this with the overall constraint~\eqref{10cond} on the chiral asymmetry and Eq.~\eqref{Vsum} this implies that 
\begin{equation}
-3 \leqslant {\rm ind}(L_a) \leqslant 0
\end{equation} 
for all $a=1,\ldots ,5$. A similar argument can be made for the $\overline{\bf 5}$--${\bf 5}$ multiplets. A positive index, ${\rm ind}(L_a\otimes L_b)>0$, would imply chiral ${\bf 5}$ multiplets in this sector. They would survive the Wilson line breaking and lead to unwanted Higgs triplets. Hence, we should require that ${\rm ind}(L_a\otimes L_b)\leq 0$ for all $a<b$ which implies that
\begin{equation}
-3 \leqslant {\rm ind}(L_a \otimes L_b) \leqslant 0 \ , \label{LLcons}
\end{equation}
for all $a<b$. 

Table~\ref{t:rk5constraint} summarizes both the consistency constraints explained earlier and the phenomenological constraints discussed in this subsection. This set of constraints will be used to classify rank five line bundle models on our $16$ Calabi-Yau manifolds. 
\begin{table}[h!t!]
{\renewcommand{\arraystretch}{1.25}
\[
\begin{array}{|c|c|}
\hline 
\mbox{~Physics~} & \mbox{~~Background geometry~~} \\\hline \hline  
\text{Gauge group} &
 c_1(V)=0 
\\ \hline

\text{Anomaly} & 
{c}_2(TX) - {c}_2(V) \in \text{ Mori}(X) 
\\ \hline

\text{Supersymmetry} & 
\mu(L_a)=0, ~\text{for }1\leq a \leq 5 
\\ \hline\hline

\text{~Three generations~} & 
{\rm ind}(V)=-3 
\\ \hline

\text{No exotics} & 
\begin{array}{l}
-3\leq {\rm ind}(L_a) \leq 0,~~\text{for }1\leq a \leq 5\, ;  \\ 
-3\leq {\rm ind}(L_a \otimes L_b) \leq 0,~~\text{for }1\leq a < b \leq 5 
\end{array} 
 
\\ \hline
\end{array}
\]}
\begin{center}
\caption{\em Consistency and phenomenological constraints imposed on rank five line bundle sums of the form~\eqref{lbs}.}
\label{t:rk5constraint}
\end{center}
\end{table}

\subsection{$SO(10)$ GUT theory}
In this case, we start with a line bundle sum~\eqref{lbs} of rank four ($n=4$) with a structure group $G=S(U(1)^4)$. The resulting four-dimensional gauge group is $H=SO(10)\times S(U(1)^4)$ and the multiplets under this gauge group which arise are
\begin{equation}
 {\bf 16}_a\;,\;\;\overline{\bf 16}_a\;,\;\;{\bf 10}_{a,b}\;,\;\;{\bf 1}_{a,b}\; .
\end{equation} 
In analogy to the $SU(5)$ case, the subscripts $a,b,\dots =1,\ldots ,4$ indicate which of the four $U(1)$ symmetries the multiplet is charged under. A ${\bf 16}_a$ ($\overline{\bf 16}_a$) multiplet carries charge $1$ ($-1$) under the $a^{\rm th}$ $U(1)$ symmetry and is uncharged under the others. A ${\bf 10}_{a,b}$ multiplet, where $a<b$, carries charge $1$ under the $a^{\rm th}$ and $b^{\rm th}$ $U(1)$ symmetry and is otherwise uncharged while a singlet ${\bf 1}_{a,b}$, where $a\neq b$, has charge $1$ under the $a^{\rm th}$ $U(1)$ and charge $-1$ under the $b^{\rm th}$ $U(1)$.

The multiplicity of each of the above multiplets is computed from associate cohomology groups as indicated in Table~\ref{t:spectrum-rk4}.
\begin{table}[h!]
{\renewcommand{\arraystretch}{1.1}
\[
\begin{array}{|c|c|c|}
\hline 
\mbox{~$SO(10)\times S(U(1)^4)$ repr.~} & \mbox{~~associated cohomology~~} & \mbox{~~contained in~~}\\
\hline \hline 
\bold{16}_a & H^{1}(X,L_a)& H^{1}(X,V) \\ \hline
\bold{\overline {16}}_a & H^1(X,L^{*}_a) &H^{1}(X,V^*) \\ \hline
\bold{10}_{a,b} & H^1(X, L_a \otimes L_b)&H^1(X,\wedge^2 V)\\ \hline
\bold{1}_{a,b} &  H^1(X,L_a \otimes L_b^*) & H^1(X,V \otimes V^{*})\\ \hline
\end{array}
\]}
\begin{center}
\caption{\em The spectrum of $SO(10)$ models and associated cohomology groups. }
\label{t:spectrum-rk4}
\end{center}
\end{table}
The three generation condition on the ${\bf 16}$--$\overline{\bf 16}$ multiplets remains the same:
\begin{equation}
{\rm ind}(V)=-3\; .
\end{equation}
For rank four bundles Eq.~\eqref{v2andv1} implies that ${\rm ind}(\wedge^2 V)=0$ so no further constraint needs to be imposed. In analogy with the $SU(5)$ case, in order to avoid $\overline{\bf 16}$ exotics, we should impose that 
\begin{equation}
 -3\leq {\rm ind}(L_a)\leq 0
\end{equation}
for all $a=1,\ldots ,4$. The line bundle indices can be explicitly computed from Eq.~\eqref{ind-linebundle}. The ${\bf 10}$ sector is automatically vector-like so no further constraint analogous to Eq.~\eqref{LLcons} is required.
 
Table~\ref{t:rk4constraint} summarizes the consistency constraints explained earlier and the phenomenological constraints discussed above. These constraints will be used to classify rank four line bundle sums on our $16$ manifolds.
\begin{table}[h!t!]
{\renewcommand{\arraystretch}{1.25}
\[
\begin{array}{|c|c|}
\hline 
\mbox{~Physics~} & \mbox{~~Background geometry~~} \\
\hline \hline 
\text{Gauge group} &
 c_1(V)=0 
\\ \hline

\text{Anomaly} & 
{\rm ch}_2(TX) - {\rm ch}_2(V) \in \text{ Mori}(X) 
\\ \hline

\text{Supersymmetry} & 
\mu(L_a)=0, ~\text{for }1\leq a \leq 4 
\\ \hline\hline

\text{~Three generations~} & 
{\rm ind}(V)=-3 
\\ \hline

\text{No exotics} & 
\begin{array}{l}
-3\leq {\rm ind}(L_a) \leq 0,~~\text{for }1\leq a \leq 4 
\end{array} 
\\ \hline
\end{array}
\]}
\begin{center}
\caption{\em Consistency and phenomenological constraints on rank four line bundles of the form~\eqref{lbs}.}
\label{t:rk4constraint}
\end{center}
\end{table}

\subsection{Search algorithm}

In principle, the scanning procedure is straight-forward now. We firstly generate all the single line bundles, $L={\cal O}_X({\bf k})$ with entries $k^r$ in a certain range and with their index between $-3$ and $0$. Then we compose these line bundles into rank four or five sums imposing the constraints detailed in Table~\ref{t:rk5constraint} and~\ref{t:rk4constraint}, respectively, as we go along and at the earliest possible stage. 

Which range of line bundle entries $k^r_a$ should we consider in this process? Unfortunately, we are not aware of a finiteness proof for line bundle sums which satisfy the constraints in Table~\ref{t:rk5constraint} and~\ref{t:rk4constraint}, nor do we know how to derive a concrete theoretical bound on the maximal size of the entries $k^r_a$ from those constraints. Lacking such a bound we proceed computationally. For a given positive integer $k_{\rm max}$ we can find all line bundle models with $k_a^r \in [-k_{\rm max}, k_{\rm max}]$. We do this for increasing values $k_{\rm max}=1,2,3,\dots$ and find the viable models for each value. If the number of these models does not increase for three consecutive $k_{\rm max}$ values, the search is considered complete. In this way, we are able to verify finiteness and find the complete set of viable models for rank five bundles. For rank four, we find the complete set for some of the manifolds but are limited by computational power for the others.

Finally, there is a practical step for simplifying the bundle search.
If the K\"ahler cone, in the form given by the original toric data, does not coincide with the positive region where all $t^r>0$ it is useful to arrange this by a suitable basis transformation. This makes checking certain properties, such as the effectiveness of a given curve class, easier. We refer to Ref.~\cite{He:2009wi} for details.

\section{Results}\label{sec4}
In this section, we describe the results of our scans for phenomenologically attractive $SU(5)$ and $SO(10)$ line bundle GUT models on the $14$ favourable Calabi-Yau three-folds out of our $16$ special ones.  

\subsection{$SU(5)$ GUT theory}
\begin{figure}[H]
\begin{center}
\includegraphics[scale=0.53]{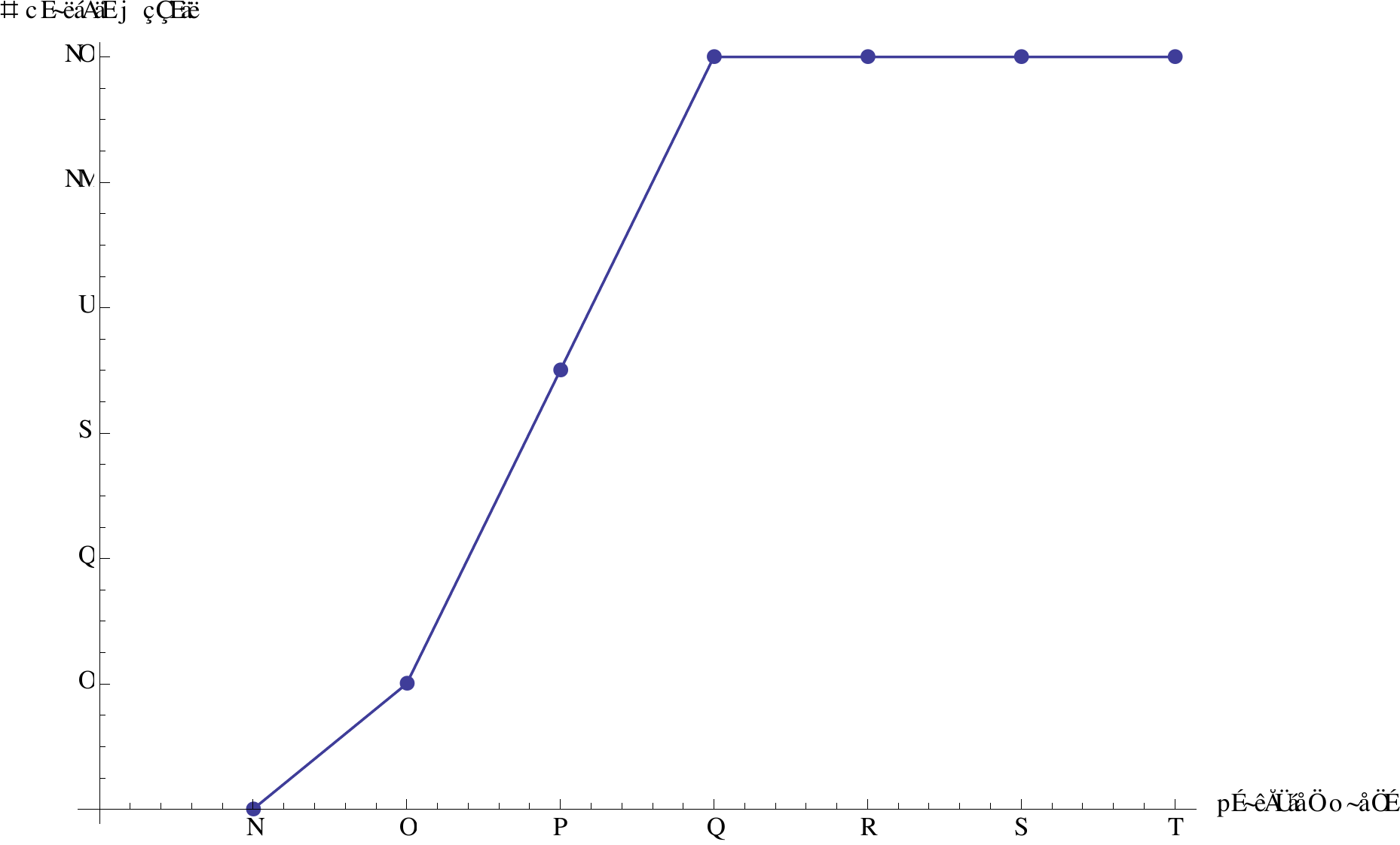}   
\end{center}
\caption{\em The number of viable line-bundle models on $X_9$ as a function of $k_{\rm max}$.}
\label{continuity}
\end{figure}

For the rank five line bundle sums we are able to verify finiteness computationally for each manifold, using the method based on scanning over entries $k_a^r$ with $-k_{\rm max}\leq k_a^r\leq k_{\rm max}$ for increasing $k_{\rm max}$, as explained above. As an illustration, we have plotted the number of viable models on $X_9$ as a function of $k_{\rm max}$ in Fig.~\ref{continuity}. As is evident from the figure, the number saturates at $k_{\rm max}=4$ and stays constant thereafter. A similar behaviour is observed for all other spaces. 
Recall from Table~\ref{t:the16} that amongst the favourable base manifolds $X_{i=1, \cdots, 14}$, only $X_1$ has Picard number 1, $X_2$ and $X_4$ have Picard number 2, $X_5$, $X_6$, $X_7$, $X_8$, $X_{14}$ have Picard number 3, and $X_3$, $X_9$, $X_{10}$, $X_{11}$, $X_{12}$, $X_{13}$ have Picard number 4.  
It turns out that viable models arise on all the six manifolds with Picard number $4$ and on two out of the five manifolds with Picard number 3, namely $X_{6}$ and $X_{14}$, in total $122$ models. The number of models for each manifold is summarized in Table~\ref{numberstable} and the explicit line bundle sums are given in Appendix~\ref{su5result}. A line bundle data set can be downloaded from Ref.~\cite{data}.
\begin{table}[h!]

\begin{center}
\begin{footnotesize}
\resizebox{\textwidth}{!}{ %
\begin{tabular}{|c||c|c|c|c|c|c|c|c|c|c|c|c|c|c||c|}\hline
&$X_1$&$X_2$&$X_3$&$X_4$&$X_5$&$X_6$&$X_7$&$X_8$&$X_9$&$X_{10}$&$X_{11}$&$X_{12}$&$X_{13}$&$X_{14}$&total\\
\hline\hline
\# $SU(5)$ &$0$&$0$&10&$0$&0&2&0&0&12&25&54&1&17&1&122\\\hline
max.~$|k_a^r|$&-&-&4&-&-&4&-&-&4&5&5&4&5&4&\\\hline\hline
\# $SO(10)$ &$0$&$0$&$7017^{\;*}$&$0$&$5$&$13$&$0$&$9$&2207&$4416^{\;*}$&$8783^{\;*}$&$1109^{\;*}$&$5283^{\;*}$&$28$&28870\\\hline
max.~$|k_a^r|$&-&-&17&-&6&7&-&4&15&20&19&21&21&7&\\\hline\hline
\end{tabular}}
\caption{\em Numbers of viable rank five ($SU(5)$) and rank four ($SO(10)$) line bundle models and maximal value of $|k_a^r|$ for each base manifold. For the $SO(10)$ cases marked with a star numbers are converging but have not quite saturated despite the large entries.}
\label{numberstable}
\end{footnotesize}
\end{center}
\end{table}

\subsection{$SO(10)$ GUT theory}

As in the $SU(5)$ cases, viable models only arise on base manifolds with Picard number greater than 2. It turns out that amongst the five Picard number 3 manifolds, $X_7$ does not admit any viable models, and the other four, $X_5$, $X_6$, $X_8$, $X_{14}$ admit 5, 13, 9, 28 bundles, respectively. For all those cases, the scan has saturated according to our criterion and the complete set of viable models has been found. In total this is 55 models which are listed in Appendix~\ref{su5result}.
For the other six manifolds $X_3$, $X_9$, $X_{10}$, $X_{11}$, $X_{12}$, $X_{13}$, all with Picard number four, only $X_9$ is complete and admits 2207 bundles. For the others, the number of viable bundles is converging but still growing slowly despite the large range of integer entries. The number of models found in each case is summarized in Table~\ref{numberstable} and the complete data sets can be downloaded from Ref.~\cite{data}. 

\subsection{An $SU(5)$ example}
To illustrate our results we would like to present explicitly one example from our data set, a three generation $SU(5)$ GUT theory on the Calabi-Yau manifold $X_9$. We recall that $X_9$ is a Picard number four manifold, constructed from eight homogeneous coordinates (see Appendix~\ref{down3} for details). From Table~\ref{numberstable} we can see that there are $12$ viable $SU(5)$ models on this manifold, with line bundle entries in the range $-4\leq k_a^r\leq 4$. 

Let us consider the first of these models from the table in Appendix~\ref{su5result} which is specified by a line bundle sum $V$ of the five line bundles
\begin{equation}
 L_1={\cal O}_X(-4, 0, 1, 1),\;\;L_2={\cal O}_X (1, 3, -1, -1),\;\;L_3=L_4=L_5={\cal O}_X(1, -1, 0, 0)\; . \label{Vex}
 \end{equation}
Evidently, $c_1(V)=0$ and, since three of the line bundles are the same, only two slope-zero conditions~\eqref{slope0} have to be satisfied in the four-dimensional K\"ahler cone. With the intersection numbers and K\"ahler cone given in Appendix~\ref{disofud}, we find that this can indeed be achieved. Further, $c_2(TX)=(12,12,12,4)$ and, from Eq.~\eqref{c2V}, $c_2(V)=(3,5,9,-7)$ so that $c_2(TX)-c_2(V)=(9,7,3,11)$ which represents a class in the Mori cone. Hence, the model can be completed to an anomaly-free model. By construction we have, of course, ${\rm ind}(V)={\rm ind}(\wedge^2V)=-3$ but, in general, the distribution of this chiral asymmetry over the various line bundle sector depends on the model. For our example, the only non-zero line bundle cohomologies are ${\rm ind}(L_1)=-3$ and ${\rm ind}(L_2\otimes L_3)={\rm ind}(L_2\otimes L_4)={\rm ind}(L_2\otimes L_5)=-1$ which implies a chiral spectrum 
\begin{equation}
 {\bf 10}_1,\; {\bf 10}_1,\; {\bf 10}_1,\; \overline{\bf 5}_{2,3},\;\overline{\bf 5}_{2,4},\;\overline{\bf 5}_{2,5}\; .
\end{equation}
Hence, the all three chiral ${\bf 10}$ multiplets are charged under the first $U(1)$ symmetry and uncharged under the others. Although, at this stage, we do not know the charge of the Higgs multiplet ${\bf 5}^{\bar{H}}$ it is clear that all up Yukawa couplings ${\bf 5}^{\bar{H}}{\bf 10}\,{\bf 10}$ are forbidden (perturbatively and at the Abelian locus). Indeed, for those terms to be $S(U(1)^5)$ invariant we require a Higgs multiplet with charge $-2$ under the first $U(1)$ and uncharged otherwise, a charge pattern which is not available at the Abelian locus. 

We also note from Eq.~\eqref{Vex} that the matrix $(k_a^r)$ of line bundle entries has rank two. This means that two of the four $U(1)$ symmetries are Green-Schwarz anomalous with corresponding super heavy gauge bosons while the other two are non-anomalous with massless gauge bosons. Those latter two $U(1)$ symmetries can be spontaneously broken, and their gauge bosons removed from the low-energy spectrum, by moving away from the line bundle locus (see Ref.~\cite{Anderson:2012yf} for details).  

\section{Conclusion and outlook}\label{sec5}

In this paper, we have studied heterotic model building on the sixteen families of torically generated Calabi-Yau three-folds with non-trivial first fundamental group~\cite{Batyrev:2005ts}. From those $16$ manifolds, we have selected the $14$ favourable three-folds and we have classified phenomenologically attractive $SU(5)$ and $SO(10)$ line bundle GUT models thereon. Concretely, we have searched for $SU(5)$ and $SO(10)$ GUT models which are supersymmetric, anomaly free and have the correct values of the chiral asymmetries to produce a three-family standard model spectrum (after subsequent inclusion of a Wilson line). For $SU(5)$ we have succeeded in finding all such line bundle models on the $14$ base spaces, thereby proving finiteness of the class computationally. The result is a total of $122$ $SU(5)$ GUT models. 

For $SO(10)$ we have obtained a complete classification for all spaces up to Picard number three, resulting in a total of $55$ $SO(10)$ GUT models. For the other six manifolds, all with Picard number four, only one ($X_9$) was amenable to a complete classification. For the other five manifolds, although the number of models were converging with increasing line bundle entries, they had not quite saturated even at fairly high values of about $k_{\rm max}=20$. We expect that we have found the vast majority of models on these manifolds with a small fraction containing some large line bundle entries still missing. Altogether we find 28870 viable $SO(10)$ models.
All models, both for $SU(5)$ and $SO(10)$, can be download from the website \cite{data}.

The main technical obstacle to determine the full spectrum of these models -- before and after Wilson line breaking -- is the computation of line bundle cohomology on torically defined Calabi-Yau manifolds. We hope to address this problem in the future.

We consider the present work as the first step in a programme of classifying all line bundle standard models on the Calabi-Yau manifolds in the Kreuzer-Skarke list. A number of technical challenges have to be overcome in order to complete this programme, including a classification of freely-acting symmetries for these Calabi-Yau manifolds and the aforementioned computation of line bundle cohomology.

\appendix

\section{Toric Data} \label{down3}
\setlongtables 
\LTcapwidth=\textwidth
{\renewcommand{\arraystretch}{1.07}
\begin{longtable}[thb]{|c||c|c|}
\hline
\endhead
\hline
\multicolumn{3}{r}{\scriptsize \slshape continued in the next page}
\endfoot
\endlastfoot
\hline 
~~~$i$~~~ & \small Vertices of $\widetilde {\Delta_i}$  &  \small Vertices of $\Delta_i$  \\  \hline \hline 
\endfirsthead 
\hline 
~~~$i$~~~ & \small Vertices of $\widetilde {\Delta_i}$  & \small  Vertices of $\Delta_i$   \\  \hline \hline 
\endhead 
%
1& 
\scriptsize $\left(\begin{array}{*{20}{c}}
{\tilde x_1}&{\tilde x_2}&{\tilde x_3}&{\tilde x_4}&{\tilde x_5} \\ \hline
4&{ - 1}&{ - 1}&{ - 1}&{ - 1}\\
{ - 1}&0&1&0&0\\
{ - 1}&1&0&0&0\\
{ - 1}&0&0&1&0
\end{array}\right)$
& 
\scriptsize$\left(\begin{array}{l}
{\begin{array}{*{20}{c}}
{{x_1}}&{{x_2}}&{{x_3}}&{{x_4}}&{{x_5}}\\ \hline
0&{ - 5}&{ 0}&{ 0}&{ 5}\\
{ -4}&1&0&3&0\\
{ -2}&0&1&1&0\\
{ 1}&-1&0&-1&1
\end{array}} 
\end{array}\right)$ \\ \hline
2&
\scriptsize$\left(\begin{array}{l}
{\begin{array}{*{24}{c}}
{{\tilde x_1}}&{{\tilde x_2}}&{{\tilde x_3}}&{{\tilde x_4}}&{{\tilde x_5}}&{{\tilde x_6}} \\ \hline
2&{ - 1}&{-1}&{ -1}&{ -1}&{2}\\
{0}&1&0&0&0&-1\\
{ 0}&0&1&0&0&-1\\
{-1}&0&0&1&0&0
\end{array}}
\end{array}\right)$
&
\scriptsize$\left(\begin{array}{l}
{\begin{array}{*{24}{c}}
{{x_1}}&{{x_2}}&{{x_3}}&{{x_4}}&{{x_5}}&{{x_6}} \\ \hline
3&{ 0}&{0}&{ 3}&{ 0}&{0}\\
{-1}&0&0&2&-1&0\\
{ 0}&1&0&1&-1&-1\\
{1}&0&1&0&-1&-1
\end{array}} 
\end{array}\right)$ \\ \hline
3&
\scriptsize$\begin{array}{l}
\left( {\begin{array}{*{32}{c}}
{{\tilde x_1}}&{{\tilde x_2}}&{{\tilde x_3}}&{{\tilde x_4}}&{{\tilde x_5}}&{{\tilde x_6}}&{{\tilde x_7}}&{{\tilde x_8}} \\ \hline
1&{ - 1}&{-1}&{ -1}&{ 1}&{1}&{1}&{-1}\\
{0}&1&0&0&0&0&-1&0\\
{ 0}&0&1&0&0&-1&0&0\\
{0}&0&0&1&-1&0&0&0
\end{array}} \right)
\end{array}$
&
\scriptsize$\begin{array}{l}
\left( {\begin{array}{*{32}{c}}
{{x_1}}&{{x_2}}&{{x_3}}&{{x_4}}&{{x_5}}&{{x_6}}&{{x_7}}&{{x_8}} \\ \hline
2&{ 0}&{0}&{ 0}&{ 0}&{0}&{0}&{-2}\\
{0}&-1&0&1&-1&0&1&0\\
{ 0}&0&-1&1&-1&1&0&0\\
{1}&0&0&1&-1&0&0&-1
\end{array}} \right)
\end{array}$\\ \hline
4&
\scriptsize$\begin{array}{l}
\left( {\begin{array}{*{24}{c}}
{{\tilde x_1}}&{{\tilde x_2}}&{{\tilde x_3}}&{{\tilde x_4}}&{{\tilde x_5}}&{{\tilde x_6}} \\ \hline
-1&{ 2}&{-1}&{ -1}&{ -1}&{-1}\\
{0}&-1&1&0&0&0\\
{3}&-1&0&0&0&1\\
{-1}&0&0&1&0&0
\end{array}} \right)
\end{array}$
&
\scriptsize$\begin{array}{l}
\left({\begin{array}{*{24}{c}}
{{x_1}}&{{x_2}}&{{x_3}}&{{x_4}}&{{x_5}}&{{x_6}} \\ \hline
3&{0}&{0}&{ 0}&{ -3}&{0}\\
{-2}&0&1&0&-1&-1\\
{-1}&1&0&0&-2&-1\\
{-2}&0&0&1&1&0
\end{array}} \right)
\end{array}$
\\ \hline
5&
\scriptsize$\begin{array}{l}
\left({\begin{array}{*{28}{c}}
{{\tilde x_1}}&{{\tilde x_2}}&{{\tilde x_3}}&{{\tilde x_4}}&{{\tilde x_5}}&{{\tilde x_6}}&{{\tilde x_7}} \\ \hline
-1&-1&1& -1&-1&-1&-1\\
4&0&-1&0&0&0&2\\
-2&2&0&0&0&1&-1\\
-1&0&0&1&0&0&0
\end{array}} \right)
\end{array}$
&
\scriptsize$\begin{array}{l}
\left( {\begin{array}{*{28}{c}}
{{x_1}}&{{x_2}}&{{x_3}}&{{x_4}}&{{x_5}}&{{x_6}}&{{x_7}} \\ \hline
-4&0&4& 0&0&2&-2\\
-1&0&2&-1&0&1&-1\\
0&1&1&-2&0&1&-1\\
-3&0&0&-1&1&0&-2
\end{array}} \right)
\end{array}$\\ \hline
6&
\scriptsize$\begin{array}{l}
\left( {\begin{array}{*{28}{c}}
{{\tilde x_1}}&{{\tilde x_2}}&{{\tilde x_3}}&{{\tilde x_4}}&{{\tilde x_5}}&{{\tilde x_6}}&{{\tilde x_7}} \\ \hline
-1&1&-1& -1&-1&-1&-1\\
2&-1&0&2&0&0&0\\
0&0&0&-1&0&1&0\\
-1&0&0&-1&2&1&1
\end{array}} \right)
\end{array}$
&
\scriptsize$\begin{array}{l}
\left( {\begin{array}{*{28}{c}}
{{x_1}}&{{x_2}}&{{x_3}}&{{x_4}}&{{x_5}}&{{x_6}}&{{x_7}} \\ \hline
-4&0&0& 0&2&0&-2\\
-3&1&0&-1&0&-2&-2\\
1&0&1&-1&0&-1&0\\
-1&0&0&-1&1&0&-1
\end{array}} \right)
\end{array}$
\\ \hline
7&
\scriptsize$\begin{array}{l}
\left( {\begin{array}{*{28}{c}}
{{\tilde x_1}}&{{\tilde x_2}}&{{\tilde x_3}}&{{\tilde x_4}}&{{\tilde x_5}}&{{\tilde x_6}}&{{\tilde x_7}} \\ \hline
-1&-1&1& -1&-1&-1&-1\\
0&2&-1&0&0&0&0\\
2&-1&0&0&0&0&1\\
-1&0&0&0&2&1&0
\end{array}} \right)
\end{array}$
&
\scriptsize$\begin{array}{l}
\left( {\begin{array}{*{28}{c}}
{{x_1}}&{{x_2}}&{{x_3}}&{{x_4}}&{{x_5}}&{{x_6}}&{{x_7}}
\end{array}} \right\}\\
\left( {\begin{array}{*{28}{c}}
-4&0&0& 0&2&-2 &0\\
-3&0&1&-1&0&-2&-1\\
-7&1&0&-1&0&-4&2\\
-1&0&0&-1&1&-1&0
\end{array}} \right)
\end{array}$\\ \hline
8&
\scriptsize$\begin{array}{l}
\left({\begin{array}{*{28}{c}}
{{\tilde x_1}}&{{\tilde x_2}}&{{\tilde x_3}}&{{\tilde x_4}}&{{\tilde x_5}}&{{\tilde x_6}}&{{\tilde x_7}} \\ \hline
-1&1&-1&-1&-1&-1&-1\\
2&-1&0&0&0&0&0\\
-1&0&2&0&2&0&1\\
0&0&0&1&-1&0&0
\end{array}} \right)
\end{array}$
&
\scriptsize$\begin{array}{l}
\left( {\begin{array}{*{28}{c}}
{{x_1}}&{{x_2}}&{{x_3}}&{{x_4}}&{{x_5}}&{{x_6}}&{{x_7}} \\ \hline
-2&0&0&4&0&4&2\\
-2&1&0&1&-1&0&0\\
-1&0&0&3&-1&2&1\\
-1&0&1&2&0&1&1
\end{array}} \right)
\end{array}$
\\ \hline
9&
\scriptsize$\begin{array}{l}
\left({\begin{array}{*{32}{c}}
{{\tilde x_1}}&{{\tilde x_2}}&{{\tilde x_3}}&{{\tilde x_4}}&{{\tilde x_5}}&{{\tilde x_6}}&{{\tilde x_7}}&{{\tilde x_8}} \\ \hline
3&-1&-1&-1&1&-1&-1&1\\
0&0&0&1&-1&0&0&0\\
-2&2&0&0&0&0&1&-1\\
-1&0&1&0&0&0&0&0
\end{array}} \right)
\end{array}$
&
\scriptsize$\begin{array}{l}
\left({\begin{array}{*{32}{c}}
{{x_1}}&{{x_2}}&{{x_3}}&{{x_4}}&{{x_5}}&{{x_6}}&{{x_7}}&{{x_8}} \\ \hline
-4&4&0&0&0&0&2&-2\\
-1&2&0&0&0&-1&1&-1\\
0&1&1&0&0&-2&1&-1\\
1&0&0&1&-1&-1&0&0
\end{array}} \right)
\end{array}$\\ \hline
10&
\scriptsize$\begin{array}{l}
\left({\begin{array}{*{32}{c}}
{{\tilde x_1}}&{{\tilde x_2}}&{{\tilde x_3}}&{{\tilde x_4}}&{{\tilde x_5}}&{{\tilde x_6}}&{{\tilde x_7}}&{{\tilde x_8}} \\ \hline
-1&1&-1&-1&1&-1&-1&-1\\
0&-1&2&0&0&0&0&1\\
2&0&0&0&-1&0&1&0\\
-1&0&0&1&0&0&0&0
\end{array}} \right)
\end{array}$
&
\scriptsize$\begin{array}{l}
\left({\begin{array}{*{32}{c}}
{{x_1}}&{{x_2}}&{{x_3}}&{{x_4}}&{{x_5}}&{{x_6}}&{{x_7}}&{{x_8}} \\ \hline
0&-4&0&0&2&0&0&-2\\
-1&1&2&-1&0&0&1&0\\
0&-1&0&-1&1&0&0&-1\\
-1&0&1&0&0&1&1&0
\end{array}} \right)
\end{array}$ \\ \hline
11&
\scriptsize$\begin{array}{l}
\left({\begin{array}{*{32}{c}}
{{\tilde x_1}}&{{\tilde x_2}}&{{\tilde x_3}}&{{\tilde x_4}}&{{\tilde x_5}}&{{\tilde x_6}}&{{\tilde x_7}}&{{\tilde x_8}} \\ \hline
1&1&1&-1&-1&-1&-1&-1\\
0&0&-1&0&0&1&0&0\\
0&-1&0&0&0&0&1&0\\
-1&0&0&0&2&0&0&1
\end{array}} \right)
\end{array}$
&
\scriptsize$\begin{array}{l}
\left({\begin{array}{*{32}{c}}
{{x_1}}&{{x_2}}&{{x_3}}&{{x_4}}&{{x_5}}&{{x_6}}&{{x_7}}&{{x_8}} \\ \hline
0&0&0&2&-2&0&0&0\\
1&-1&0&0&0&0&-1&-1\\
0&1&-1&0&0&1&-1&0\\
0&1&0&1&-1&0&-1&0
\end{array}} \right)
\end{array}$\\ \hline
12&
\scriptsize$\begin{array}{l}
\left({\begin{array}{*{32}{c}}
{{\tilde x_1}}&{{\tilde x_2}}&{{\tilde x_3}}&{{\tilde x_4}}&{{\tilde x_5}}&{{\tilde x_6}}&{{\tilde x_7}}&{{\tilde x_8}} \\ \hline
-1&1&1&-1&-1&-1&-1&-1\\
0&0&-1&0&0&1&0&0\\
2&-1&0&0&0&0&0&1\\
-1&0&0&0&2&0&1&0
\end{array}} \right)
\end{array}$
&
\scriptsize$\begin{array}{l}
\left({\begin{array}{*{32}{c}}
{{x_1}}&{{x_2}}&{{x_3}}&{{x_4}}&{{x_5}}&{{x_6}}&{{x_7}}&{{x_8}} \\ \hline
0&0&-2&0&0&2&0&0\\
0&1&0&-1&-3&0&-2&-1\\
1&0&0&-1&-1&0&-1&0\\
0&0&-1&-1&1&1&0&0
\end{array}} \right)
\end{array}$
\\ \hline
13&
\scriptsize$\begin{array}{l}
\left({\begin{array}{*{32}{c}}
{{\tilde x_1}}&{{\tilde x_2}}&{{\tilde x_3}}&{{\tilde x_4}}&{{\tilde x_5}}&{{\tilde x_6}}&{{\tilde x_7}}&{{\tilde x_8}} \\ \hline
1&-1&-1&-1&-1&1&-1&-1\\
0&0&0&1&0&-1&0&0\\
-1&2&0&0&2&0&0&1\\
0&0&1&0&-1&0&0&0
\end{array}} \right)
\end{array}$
&
\scriptsize$\begin{array}{l}
\left({\begin{array}{*{32}{c}}
{{x_1}}&{{x_2}}&{{x_3}}&{{x_4}}&{{x_5}}&{{x_6}}&{{x_7}}&{{x_8}} \\ \hline
0&0&0&-2&2&0&0&0\\
-1&-1&2&0&0&0&3&1\\
0&-1&0&-1&1&0&1&0\\
-1&0&1&0&0&1&2&1
\end{array}} \right)
\end{array}$\\ \hline
14&
\scriptsize$\begin{array}{l}
\left({\begin{array}{*{28}{c}}
{{\tilde x_1}}&{{\tilde x_2}}&{{\tilde x_3}}&{{\tilde x_4}}&{{\tilde x_5}}&{{\tilde x_6}}&{{\tilde x_7}} \\ \hline
1&-1&-1&-1&-1&-1&-1\\
-1&2&0&2&0&2&0\\
0&-1&1&0&0&-1&1\\
0&0&1&0&0&-1&0
\end{array}} \right)
\end{array}$
&
\scriptsize$\begin{array}{l}
\left({\begin{array}{*{28}{c}}
{{x_1}}&{{x_2}}&{{x_3}}&{{x_4}}&{{x_5}}&{{x_6}}&{{x_7}} \\ \hline
0&0&0&-2&2&0&0\\
-1&-1&2&2&0&0&3\\
0&-1&0&-1&1&0&1\\
-1&0&1&2&0&1&2
\end{array}} \right)
\end{array}$
\\ \hline
15&
\scriptsize$\begin{array}{l}
\left({\begin{array}{*{32}{c}}
{{\tilde x_1}}&{{\tilde x_2}}&{{\tilde x_3}}&{{\tilde x_4}}&{{\tilde x_5}}&{{\tilde x_6}}&{{\tilde x_7}}&{{\tilde x_8}} \\ \hline
-1&3&-1&-1&-1&-1&-1&1\\
2&-2&0&0&0&0&1&-1\\
0&-1&1&0&0&0&0&0\\
-1&0&0&0&2&1&0&0
\end{array}} \right)
\end{array}$
&
\scriptsize$\begin{array}{l}
\left({\begin{array}{*{32}{c}}
{{x_1}}&{{x_2}}&{{x_3}}&{{x_4}}&{{x_5}}&{{x_6}}&{{x_7}}&{{x_8}} \\ \hline
-4&0&4&-4&0&-4&-2&2\\
-1&0&2&-3&0&-2&-1&1\\
-2&1&1&-2&0&-2&-1&1\\
-1&0&0&-1&1&-1&0&0
\end{array}} \right)
\end{array}$\\ \hline
16&
\scriptsize$\begin{array}{l}
\left({\begin{array}{*{32}{c}}
{{\tilde x_1}}&{{\tilde x_2}}&{{\tilde x_3}}&{{\tilde x_4}}&{{\tilde x_5}}&{{\tilde x_6}}&{{\tilde x_7}}&{{\tilde x_8}} \\ \hline
-3&-1&-1&-1&-1&-1&-1&1\\
-1&1&0&0&0&0&0&0\\
-2&0&0&2&0&2&1&-1\\
0&0&1&0&0&-1&0&0
\end{array}} \right)
\end{array}$
&
\scriptsize$\begin{array}{l}
\left({\begin{array}{*{32}{c}}
{{x_1}}&{{x_2}}&{{x_3}}&{{x_4}}&{{x_5}}&{{x_6}}&{{x_7}}&{{x_8}} \\ \hline
4&0&0&0&-4&-4&2&-2\\
2&0&0&1&-3&-2&1&-1\\
1&1&0&0&-2&-2&1&-1\\
0&0&1&1&-1&-1&0&0
\end{array}} \right)
\end{array}$ \\\hline

\caption{{\em The sixteen pairs $(\widetilde {\Delta_i}, \Delta_i)$ of reflexive four-polytopes, for $i=1, \cdots, 16$, each pair leading to the upstairs Calabi-Yau geometry $\widetilde{X_i} \subset \widetilde{\mathcal A_i}$ and the downstairs geometry $X_i \subset \mathcal A_i$ with $\pi_1(X_i) \neq \emptyset$. The polytopes are described in terms of their integral vertices.}}
\label{16families}
\end{longtable}
}

\section{Base Geometries: Upstairs and Downstairs} \label{disofud}

In this Appendix, we analyse the quotient relationship between the $16$ upstairs manifolds $\widetilde {X_i} \subset \widetilde{\mathcal A_i}$ and the corresponding $16$ downstairs manifolds $X_i \subset \mathcal A_i$ whose defining polytopes were given in the previous Appendix.
In addition, some geometrical properties of these manifolds relevant to model building will also be discussed. 

\subsection{An Illustrative Example: the Quintic three-fold} \label{monoexample}
Amongst the sixteen pairs is the quintic manifold $\widetilde{X_1}$ and its $\mathbb Z_5$ quotient $X_1$, which we take as an illustrative example. The corresponding two polytopes $\widetilde{\Delta_1}$ and $\Delta_1$ have $5$ vertices each. 

Firstly, the vertices of $\widetilde{\Delta_1}$ for the quintic three-fold $\widetilde{X_1}$ can be read off from Table~\ref{16families}: 
\begin{equation}
\left(\begin{array}{*{20}{c}}
{\tilde x_1}&{\tilde x_2}&{\tilde x_3}&{\tilde x_4}&{\tilde x_5} \\ \hline
4&{ - 1}&{ - 1}&{ - 1}&{ - 1}\\
{ - 1}&0&1&0&0\\
{ - 1}&1&0&0&0\\
{ - 1}&0&0&1&0
\end{array}\right) \ , 
\end{equation} 
where $\tilde x_{\rho=1, \cdots, 5}$ are the homogeneous coordinates on the ambient space $\mathbb P^4$. 
The polytope $\widetilde{\Delta_1}$ naturally leads to the usual $126$ quintic monomials in $\tilde x_\rho$; these generate the defining polynomial of the quintic Calabi-Yau three-fold $\widetilde{X_1}$.

Similarly, the vertices of $\Delta_1$ for the quotiented quintic $X_1=\widetilde{X_1}/\mathbb Z_5$ are given as follows:
\begin{equation}
\left(\begin{array}{l}
{\begin{array}{*{20}{c}}
{{x_1}}&{{x_2}}&{{x_3}}&{{x_4}}&{{x_5}}\\ \hline
0&{ - 5}&{ 0}&{ 0}&{ 5}\\
{ -4}&1&0&3&0\\
{ -2}&0&1&1&0\\
{ 1}&-1&0&-1&1
\end{array}} 
\end{array}\right) \ ,
\end{equation}
where $x_{\rho=1, \cdots, 5}$ are again the homogeneous coordinates on the corresponding toric ambient space. 
As for the generators of the defining polynomial, the polytope $\Delta_1$ leads to the following 26 monomials in $x_\rho$: 
\begin{equation}\label{laurant}
\begin{array}{l} 
~~~~~~x_2^5~,~~{x_1}x_2^3{x_3}~,~~x_2^2x_3^2{x_5}~,~~x_2^3{x_4}{x_5}~,~~{x_1}x_2^2x_5^2~,~~{x_2}{x_3}x_5^3~,~~x_5^5~,~~x_3^5~,~~x_1^2{x_2}x_3^2\ ,\\
{x_2}x_3^3{x_4}~,~~{x_1}x_3^3{x_5}~,~~x_1^2x_2^2{x_4}~,~~x_1^3{x_2}{x_5}~,~~x_2^2{x_3}x_4^2~,~~{x_1}{x_2}{x_3}{x_4}{x_5}~,~~x_1^2{x_3}x_5^2~,~~x_3^2{x_4}x_5^2\ ,\\
~~~{x_2}x_4^2x_5^2~,~~{x_1}{x_4}x_5^3~,~~x_1^5~,~~x_1^3{x_3}{x_4}~,~~{x_1}x_3^2x_4^2~,~~{x_1}{x_2}x_4^3~,~~x_1^2x_4^2{x_5}~,~~{x_3}x_4^3{x_5}~,~~x_4^5 \ .
\end{array}
\end{equation}
Now, by demanding that the $26$ monomials be invariant, we find the following phase rotation rule 
\begin{equation}\label{phase}
\{ {{\tilde x_1}} \to {x_1},~{{\tilde x_2}} \to {e^{\frac{{2i\pi }}{5}}}{x_2},~{{\tilde x_3}} \to {e^{\frac{{4i\pi }}{5}}}{x_3},~ {{\tilde x_4}} \to {e^{\frac{{6i\pi }}{5}}}{x_4},~ {{\tilde x_5}} \to {e^{\frac{{8i\pi }}{5}}}{x_5}\} \ ,
\end{equation}
which links the two sets of homogeneous coordinates. 

This phase rotation relates the two manifolds $\widetilde{X_1}$ and $X_1$ tightly. Not only the Laurant polynomials are explicitly connected, it turns out that the integral cohomology groups are also very much similar under the phase rotation.

As an example illustrating the precise relation between upstairs and downstairs space, consider  one of the 126 monomials, ${\widetilde{x}_1}\widetilde{x}_2^3{\widetilde{x}_3}$,  defining the upstairs ambient space of the quintic $\widetilde{{X}_1}$. 
If we transform this monomial using the rules in Eq.~\eqref{phase} we obtain ${\widetilde{x}_1}\widetilde{x}_2^3{\widetilde{x}_3}\rightarrow x_1  ({e^{\frac{{2i\pi }}{5}}}{x_2})^3 {e^{\frac{{4i\pi }}{5}}}{x_3} = {x_1}x_2^3{x_3}$. The phase independence of the result means that this is one of the $26$ monomials which define the downstairs manifold $X_1=\widetilde{{X}_1}/\mathbb{Z}_5$. The remaining 25 downstairs monomials can be obtained by applying this procedure systematically to all upstairs monomials~\footnote{In some cases, an additional permutation of the downstairs homogeneous coordinate has to be included, as in some of the examples in Table.~\ref{BaseGeometries}. This is to ensure that the linear relationships between divisors and integral basis are literally the same for both the upstairs and the downstairs manifolds.}.

We next turn to some relevant base geometries, most of which can be easily extracted from PALP~\cite{PALP}. 
Let us start from upstairs. 
Firstly, the Picard group of $\widetilde{X_1}$ is generated by a single element $\tilde J_1$ and all the toric divisors are rationally equivalent to $\tilde J_1$:
$${\tilde D_1} =\tilde J_1, ~{\tilde D_2} = \tilde J_1, ~{\tilde D_3} = \tilde J_1,~{\tilde D_4} =\tilde J_1,~ {\tilde D_5} = {\tilde J_1} \ .$$ 
Note that we do not carefully distinguish harmonic $(1,1)$-forms from divisors unless ambiguities arise. 
The intersection polynomial is: $$5 {\tilde J_{1}}^{3} \ , $$ which means that $d_{111}(\widetilde{X_1})=5$. 
In general, the coefficient of the monomial term $\tilde J_r \tilde J_s \tilde J_t$ in the intersection polynomial is the value of $d_{rst}(\widetilde X)$, without any symmetry factors. 
Finally, the Hodge numbers are:
$$\begin{array}{*{20}{c}}
{{h^{1,1}(\widetilde{X_1})} = 1,}&{{h^{1,2}(\widetilde{X_1})} = 101} \ ,
\end{array}$$
leading to the Euler character $\chi(\widetilde{X_1})= -200$.

As for the downstairs manifold $X_1$, the $\mathbb Z_5$-quotient of the quintic $\widetilde{X_1}$, the Picard group is again spanned by a single element $J_1$ and the toric divisors are all equivalent: $${D_1} =J_1, ~{D_2} = J_1, ~{D_3} = J_1,~{D_4} =J_1,~ {D_5} = {J_1} \ . $$
The intersection polynomial is given as: $${J_1^{3}} \ , $$
and hence, $d_{111}(X_1)=1$. 
Finally, the Hodge numbers are:
$$\begin{array}{*{20}{c}}
{{h^{1,1}(X_1)} = 1,}&{{h^{1,2}(X_1)} = 21}
\end{array}$$
and the Euler character $\chi(X_1)  =  - 40$.

Note that the intersection polynomial of $X_1$ is equal to that of $\widetilde{X_1}$ divided by $5$, the order of the discrete group $\mathbb Z_5$. This remains true for all the fourteen favorable manifolds $X_{i=1, \cdots, 14}$ in an appropriate basis of $H^{1,1}$.

\subsection{Summary of the Base Geometries}\label{sumBG}
For the remaining fifteen cases, the phase rotations of the homogeneous coordinates are not as straight-forward as in the quintic example. 
One needs to make use of some combinatorial tricks to figure out the explicit results. In some cases, permutations are also required to make the upstairs and the downstairs intersection polynomials proportional to each other. In Table~\ref{BaseGeometries}, we summarise the complete results for all the sixteen pairs of geometries. For each pair, we first present the phase rotation map between upstairs and downstairs coordinates (and the permutation of the coordinates if required). The base geometries of $\widetilde{X_i}$ and $X_i$ then follow in order: number of generating monomials,\footnote{For simplicity, we do not attempt to explicitly show the generating monomials and only give the number of viable terms. However, the idea should be clear from the quintic example in section~\ref{monoexample}.} toric divisors in terms of the $(1,1)$-form basis elements, intersection polynomial. The Hodge numbers $h^{1,1}$ and $h^{2,1}$, as well as the Euler character $\chi$ of the manifold $X$ are presented using the notation $\left[X\right]^{h^{1,1}, h^{2,1}}_{\chi\,.}$. In addition, the second Chern class $c_2(TX)$ and K\"ahler cone matrix $K$ for the downstairs manifolds are also being listed. The K\"ahler cone is then given by all K\"ahler parameters satisfying $K_{rs}t^s\geq 0$ for all $r$. 
\setlongtables 
\LTcapwidth=\textwidth
{\renewcommand{\arraystretch}{1.2}
\begin{longtable}[thb]{|c|c|}
\hline
\endhead
\hline
\endfoot
\endlastfoot
\hline 
\multicolumn{2}{c}{\scriptsize  \slshape Base Geometries: Upstairs and Downstairs} \\ \hline \hline
\endfirsthead 
\hline 
\multicolumn{2}{c}{\scriptsize  \slshape Base Geometries: Upstairs and Downstairs} \\ \hline \hline
\endhead 
\multicolumn{2}{c}{} \\
\multicolumn{2}{c}{ Pair $1$: $\{ {{\tilde x_1}} \to {x_1},~{{\tilde x_2}} \to {e^{\frac{{2i\pi }}{5}}}{x_2},~{{\tilde x_3}} \to {e^{\frac{{4i\pi }}{5}}}{x_3},~ {{\tilde x_4}} \to {e^{\frac{{6i\pi }}{5}}}{x_4},~ {{\tilde x_5}} \to {e^{\frac{{8i\pi }}{5}}}{x_5}\}$} \\
\hline \hline
$~~~~~~~~~~~~~~~~~~~~~~~~~~~~~[\widetilde{ X_1}]^{1,101}_{-200}$~~~~~~~~~~~~~~~~~~~~~~~~~~~& \scriptsize\#(monomials) = $126$ \\ \hline
\multicolumn{2}{|c|}{\scriptsize${\tilde D_1} =\tilde J_1, ~{\tilde D_2} = \tilde J_1, ~{\tilde D_3} = \tilde J_1,~{\tilde D_4} =\tilde J_1,~ {\tilde D_5} = {\tilde J_1}$} \\ \hline
\multicolumn{2}{|c|}{\scriptsize$5 {\tilde J_{1}}^{3} $}\\ \hline \hline
$[{X_1}]^{1,21}_{-40}$ & \scriptsize\#(monomials) = $26$ \\ \hline 
\multicolumn{2}{|c|}{\scriptsize$D_1=J_1, D_2=J_2, D_3=J_2, D_4=J_1, D_5=J_1, D_6=J_2$} \\ \hline
\multicolumn{2}{|c|}{\scriptsize$ { J_{1}}^{3} $}\\ \hline
{\scriptsize $c_2(TX)=(10)$} & \scriptsize $K = (1) $ \\
\hline \hline

\multicolumn{2}{c}{} \\
\multicolumn{2}{c}{Pair $2$: $\{ {{\tilde x_1}} \to {x_1},~{{\tilde x_4}} \to {e^{\frac{{2i\pi }}{3}}}{x_4},~{{\tilde x_5}} \to {e^{\frac{{4i\pi }}{3}}}{x_5},~{{\tilde x_2}} \to {x_2},~{{\tilde x_3}} \to {e^{\frac{{2i\pi }}{3}}}{x_3},~{{\tilde x_6}} \to {e^{\frac{{4i\pi }}{3}}}{x_6}\} $} \\
\hline \hline
$~~~~~~~~~~~~~~~~~~~~[\widetilde{X_2}]^{2,83}_{-162}$~~~~~~~~~~~~~~~~~~~~& \scriptsize\#(monomials) = $100$ \\ \hline
\multicolumn{2}{|c|}{\scriptsize$\tilde D_1=\tilde J_1, \tilde D_2=\tilde J_2, \tilde D_3=\tilde J_2, \tilde D_4=\tilde J_1, \tilde D_5=\tilde J_1, \tilde D_6=\tilde J_2$} \\ \hline
\multicolumn{2}{|c|}{\scriptsize$ 3\, \tilde J_1^2 \, \tilde J_2+3 \, \tilde J_1\, \tilde J_2^2$}\\ \hline \hline
$[{X_2}]^{2,29}_{-54}$ & \scriptsize\#(monomials) = $34$ \\ \hline 
\multicolumn{2}{|c|}{\scriptsize${D_1} =J_1, ~{D_2} = J_1, ~{D_3} = J_1,~{D_4} =J_1,~ {D_5} = {J_1} $} \\ \hline
\multicolumn{2}{|c|}{\scriptsize$J_1^2 \, J_2+ J_1\, J_2^2$}\\ \hline
{\scriptsize $c_2(TX) = (12,12)$} & \scriptsize $K = 
\left( {\begin{array}{*{20}{c}}
  0&1 \\ 
  1&0 
\end{array}} \right)$ \\ 
\hline \hline

\multicolumn{2}{c}{} \\
\multicolumn{2}{c}{Pair $3$: $\{ {{\tilde x_1}} \to {e^{i\pi }}{x_1},~{{\tilde x_2}} \to {e^{i\pi }}{x_2},~{{\tilde x_3}} \to {e^{i\pi }}{x_3},~{{\tilde x_4}} \to {e^{i\pi }}{x_4},~{{\tilde x_5}} \to {x_5},~{{\tilde x_6}} \to {x_6},~{{\tilde x_7}} \to {x_7},~{{\tilde x_8}} \to {x_8}\}$} \\
\hline \hline
$~~~~~~~~~~~~~~~~~~~~[\widetilde{X_3}]^{4,68}_{-128}$~~~~~~~~~~~~~~~~~~~~& \scriptsize\#(monomials) = $81$ \\ \hline
\multicolumn{2}{|c|}{\scriptsize$\tilde D_1=\tilde J_4,\tilde D_2=\tilde J_3,\tilde D_3=\tilde J_2,\tilde D_4=\tilde J_1,\tilde D_5=\tilde J_1,\tilde D_6=\tilde J_2,\tilde D_7=\tilde J_3,\tilde D_8=\tilde J_4$} \\ \hline
\multicolumn{2}{|c|}{\scriptsize$2\, \tilde J_1\, \tilde J_2\, \tilde J_3+2\, \tilde J_1\, \tilde J_2\, \tilde J_4+2\, \tilde J_1\, \tilde J_3\, \tilde J_4+2\, \tilde J_2\,\tilde  J_3\, \tilde J_4$}\\ \hline \hline
$[{X_3}]^{4,36}_{-64}$ & \scriptsize\#(monomials) = $26$ \\ \hline 
\multicolumn{2}{|c|}{\scriptsize$D_1=J_4, D_2=J_3, D_3=J_2, D_4=J_1, D_5=J_1, D_6=J_2, D_7=J_3, D_8=J_4$} \\ \hline
\multicolumn{2}{|c|}{\scriptsize$J_1\, J_2\, J_3+J_1\, J_2\, J_4+J_1\, J_3\, J_4+J_2\, J_3\, J_4$}\\ \hline
{\scriptsize $c_2(TX) = (12,12,12,12)$} & \scriptsize 
$K = 
\left( {\begin{array}{*{20}{c}}
  1&0&0&0 \\ 
  0&0&0&1 \\ 
  0&1&0&0 \\ 
  0&0&1&0 
\end{array}} \right)$ \\ 
\hline \hline

\multicolumn{2}{c}{} \\
\multicolumn{2}{c}{Pair $4$: $\{ {{\tilde x_1}} \to {x_1},~{{\tilde x_2}} \to {e^{\frac{{2i\pi }}{3}}}{x_2},~{{\tilde x_4}} \to {e^{\frac{{4i\pi }}{3}}}{x_4},~{{\tilde x_3}} \to {x_3},~{{\tilde x_5}} \to {e^{\frac{{2i\pi }}{3}}}{x_5},~{{\tilde x_6}} \to {e^{\frac{{4i\pi }}{3}}}{x_6}\}$} \\
\hline \hline
$~~~~~~~~~~~~~~~~~~~~[\widetilde{X_4}]^{4,112}_{-216}$~~~~~~~~~~~~~~~~~~~~& \scriptsize\#(monomials) = $145$ \\ \hline
\multicolumn{2}{|c|}{\scriptsize$\tilde D_1=\tilde J_1, \tilde D_2=3\, \tilde J_1+\tilde J_2, \tilde D_3=3\, \tilde J_1+\tilde J_2, \tilde D_4=\tilde J_1, \tilde D_5=\tilde J_1, \tilde D_6=\tilde J_2$} \\ \hline
\multicolumn{2}{|c|}{\scriptsize$3\, \tilde J_1^2\, \tilde J_2-9\, \tilde J_1\, \tilde J_2^2+27\, \tilde J_2^3$}\\ \hline \hline
$[{X_4}]^{2,38}_{-72}$ & \scriptsize\#(monomials) = $49$ \\ \hline 
\multicolumn{2}{|c|}{\scriptsize$D_1=J_1, D_2=3\, J1+J2, d3=3\, J1+J2, D_4=J_1, D_5=J_1, D_6=J_2$} \\ \hline
\multicolumn{2}{|c|}{\scriptsize$J_1^2\, J_2-3\, J_1\, J_2^2+9\, J_2^3$}\\ \hline
{\scriptsize $c_2(TX) = (12,-6)$} & \scriptsize 
$ K= \left( {\begin{array}{*{20}{c}}
  0&1 \\ 
  1&{ - 3} 
\end{array}} \right)
$ \\  
\hline \hline

\multicolumn{2}{c}{} \\
\multicolumn{2}{c}{Pair $5$: $\{ {{\tilde x_1}} \to {e^{i\pi }}{x_1},~{{\tilde x_2}} \to {e^{i\pi }}{x_2},~{{\tilde x_3}} \to {e^{i\pi }}{x_3},~{{\tilde x_6}} \to {e^{i\pi }}{x_6},~{{\tilde x_4}} \to {x_4},{{\tilde x_5}} \to {x_5}\}$} \\ 
\multicolumn{2}{c}{$\{ {x_3} \to {x_5},~{x_5} \to {x_3}\}$} \\
\hline \hline
$~~~~~~~~~~~~~~~~~~~~[\widetilde{X_5}]^{3,83}_{-160}$~~~~~~~~~~~~~~~~~~~~& \scriptsize\#(monomials) = $105$ \\ \hline
\multicolumn{2}{|c|}{\scriptsize$\tilde D_1=\tilde J_1, \tilde D_2=\tilde J_2, \tilde D_3=4\, \tilde J_1+2\, \tilde J_3, \tilde D_4=\tilde J_1, \tilde D_5=\tilde J_2, \tilde D_6=2\, \tilde J_1-2\, \tilde J_2+\tilde J_3, \tilde D_7=\tilde J_3$} \\ \hline
\multicolumn{2}{|c|}{\scriptsize$2\, \tilde J_1\, \tilde J_2\, \tilde J_3+4\, \tilde J_1\, \tilde J_3^2-4\, \tilde J_2\, \tilde J_3^2-16\, \tilde J_3^3$}\\ \hline \hline
$[{X_5}]^{3,43}_{-80}$ & \scriptsize\#(monomials) = $53$ \\ \hline 
\multicolumn{2}{|c|}{\scriptsize$D_1=J_2, D_2=J_1, D_3=J_1, D_4=J_2, D_5=4\, J_2+2\, J_3, D_6=-2\, J_1+2\, J_2+J_3, D_7=J_3$} \\ \hline
\multicolumn{2}{|c|}{\scriptsize$J_1\, J_2\, J_3-2\, J_1\, J_3^2+2\, J_2\, J_3^2-8\, J_3^3$}\\ \hline
{\scriptsize $c_2(TX) = (12,12,4)$}  & \scriptsize 
$K= 
\left( {\begin{array}{*{20}{c}}
  0&1&{ - 2} \\ 
  1&0&0 \\ 
  0&0&1 
\end{array}} \right)$ \\  
\hline \hline

\multicolumn{2}{c}{} \\
\multicolumn{2}{c}{Pair $6$: $\{ {{\tilde x_1}} \to {e^{i\pi }}{x_1},~{{\tilde x_2}} \to {e^{i\pi }}{x_2},~{{\tilde x_3}} \to {e^{i\pi }}{x_3},~{{\tilde x_4}} \to {e^{i\pi }}{x_4},~{{\tilde x_5}} \to {x_5},~{{\tilde x_6}} \to {x_6}\}$} \\
\multicolumn{2}{c}{$\{ {x_1} \to {x_5},{x_5} \to {x_1},{x_3} \to {x_4},{x_4} \to {x_3}\}$}\\ 
\hline \hline
$~~~~~~~~~~~~~~~~~~~~[\widetilde{X_6}]^{3,115}_{-224}$~~~~~~~~~~~~~~~~~~~~& \scriptsize\#(monomials) = $153$ \\ \hline
\multicolumn{2}{|c|}{\scriptsize$\tilde D_1=2\, \tilde J_1+\tilde J_3, \tilde D_2=4\, \tilde J_1+2\, \tilde J_2+2\, \tilde J_3, \tilde D_3=\tilde J_1, \tilde D_4=\tilde J_2, \tilde D_5=\tilde J_1, \tilde D_6=\tilde J_2, \tilde D_7=\tilde J_3$} \\ \hline
\multicolumn{2}{|c|}{\scriptsize$2\, \tilde J_1\, \tilde J_2\, \tilde J_3-4\, \tilde J_2\, \tilde J_3^2$}\\ \hline \hline
$[{X_6}]^{3,59}_{-112}$ & \scriptsize\#(monomials) = $77$ \\ \hline 
\multicolumn{2}{|c|}{\scriptsize$D_1=J_1, D_2=4\, J_1+2\, J_2+2\, J_3, D_3=J_2, D_4=J_1, D_5=2\, J_1+J_3, D_6=J_2, D_7=J_3$}\\ \hline
\multicolumn{2}{|c|}{\scriptsize$J_1\, J_2\, J_3-2\, J_2\, J_3^2$}\\ \hline
\multicolumn{2}{|c|}{\scriptsize $c_2(TX) = (12,12,0)$}\\ \hline
\multicolumn{2}{|c|}{\scriptsize 
$K_1 = \left( {\begin{array}{*{20}{c}}
  0&1&{ - 2} \\ 
  1&0&0 \\ 
  0&0&1 
\end{array}} \right)$ ,
$K_2 = \left( {\begin{array}{*{20}{c}}
  0&0&1 \\ 
  1&0&{ - 2} \\ 
  0&1&{ - 1} 
\end{array}} \right)$,
$K_{join} = \left( {\begin{array}{*{20}{c}}
  1&0&0 \\ 
  0&1&0 \\ 
  0&0&1 
\end{array}} \right)$
}\\
\hline \hline

\multicolumn{2}{c}{} \\
\multicolumn{2}{c}{Pair $7$: $\{ {\tilde x_1} \to {e^{i\pi }}{x_1},~{\tilde x_2} \to {e^{i\pi }}{x_2},~{\tilde x_3} \to {e^{i\pi }}{x_3},~{\tilde x_4} \to {e^{i\pi }}{x_4}\} $} \\
\multicolumn{2}{c}{$\{ {x_1} \to {x_5},{x_5} \to {x_1},{x_2} \to {x_3},{x_3} \to {x_2}\} $} \\ 
\hline \hline
$~~~~~~~~~~~~~~~~~~~~[\widetilde{X_7}]^{4,148}_{-288}$~~~~~~~~~~~~~~~~~~~~& \scriptsize\#(monomials) = $126$ \\ \hline
\multicolumn{2}{|c|}{\scriptsize$\tilde D_1=2\, \tilde J_1+\tilde J_2, \tilde D_2=4\, \tilde J_1+2\, \tilde J_2+\tilde J_3, \tilde D_3=8\, \tilde J_1+4\, \tilde J_2+2\, \tilde J_3, \tilde D_4=\tilde J_1, \tilde D_5=\tilde J_1, \tilde D_6=\tilde J_2, \tilde D_7=\tilde J_3$} \\ \hline
\multicolumn{2}{|c|}{\scriptsize$2\, \tilde J_1\, \tilde J_2\, \tilde J_3-4\, \tilde J_2^2\, \tilde J_3-4\, \tilde J_1\, \tilde J_3^2+16\, \tilde J_3^3$}\\ \hline \hline
$[{X_7}]^{3,75}_{-144}$ & \scriptsize\#(monomials) = $26$ \\ \hline 
\multicolumn{2}{|c|}{\scriptsize$D_1=J_1, D_2=8\, J_1+4\, J_2+2\, J_3, D_3=4\, J_1+2\, J_2+J_3, D_4=J_1, D_5=2\, J_1+J_2, D_6=J_2, D_7=J_3$} \\ \hline
\multicolumn{2}{|c|}{\scriptsize$J_1\, J_2\, J_3-2\, J_2^2\, J_3-2\, J_1\, J_3^2+8\, J_3^3$}\\ \hline
{\scriptsize $c_2(TX) = (12,0,-4)$} & \scriptsize $K = 
\left( {\begin{array}{*{20}{c}}
  1&{ - 2}&0 \\ 
  0&0&1 \\ 
  0&1&{ - 2} 
\end{array}} \right)$ \\ 
\hline \hline

\multicolumn{2}{c}{} \\
\multicolumn{2}{c}{Pair $8$: $\{ {\tilde x_1} \to {e^{i\pi }}{x_1},~{\tilde x_2} \to {e^{i\pi }}{x_2},~{\tilde x_3} \to {e^{i\pi }}{x_3},~{\tilde x_4} \to {e^{i\pi }}{x_4}\} $} \\
\hline \hline
$~~~~~~~~~~~~~~~~~~~~[\widetilde{X_8}]^{4,148}_{-288}$~~~~~~~~~~~~~~~~~~~~& \scriptsize\#(monomials) = $201$ \\ \hline
\multicolumn{2}{|c|}{\scriptsize$\tilde D_1=2\, \tilde J_1+2\, \tilde J_2+\tilde J_3, \tilde D_2=4\, \tilde J_1+4\, \tilde J_2+2\, \tilde J_3, \tilde D_3=\tilde J_2, \tilde D_4=\tilde J_1, \tilde D_5=\tilde J_1, \tilde D_6=\tilde J_2, \tilde D_7=\tilde J_3$} \\ \hline
\multicolumn{2}{|c|}{\scriptsize$2\, \tilde J_1\, \tilde J_2\, \tilde J_3-4\, \tilde J_1\, \tilde J_3^2-4\, \tilde J_2\, \tilde J_3^2+16\, \tilde J_3^3$}\\ \hline \hline
$[{X_8}]^{3,75}_{-144}$ & \scriptsize\#(monomials) = $101$ \\ \hline 
\multicolumn{2}{|c|}{\scriptsize$D_1=2\, J_1+2\, J_2+J_3, D_2=4\, J_1+4\, J_2+2\, J_3, D_3=J_2, D_4=J_1, D_5=J_1, D_6=J_2, D_7=J_3$} \\ \hline
\multicolumn{2}{|c|}{\scriptsize$J_1\, J_2\, J_3-2\, J_1\, J_3^2-2\, J_2\, J_3^2+8\, J_3^3$}\\ \hline
{\scriptsize $c_2(TX) = (12,12,-4)$} & \scriptsize $ K = \left( {\begin{array}{*{20}{c}}
  0&0&1 \\ 
  1&0&{ - 2} \\ 
  0&1&{ - 2} 
\end{array}} \right)$ \\ 
\hline \hline

\multicolumn{2}{c}{} \\
\multicolumn{2}{c}{Pair $9$: $\{ {\tilde x_1} \to {e^{i\pi }}{x_1},~{\tilde x_2} \to {e^{i\pi }}{x_2},~{\tilde x_4} \to {e^{i\pi }}{x_4},~{\tilde x_7} \to {e^{i\pi }}{x_7}\} $} \\
\multicolumn{2}{c}{$\{ {x_3} \to {x_6},~{x_6} \to {x_3}\} $}\\
\hline \hline
$~~~~~~~~~~~~~~~~~~~~[\widetilde{X_9}]^{4,52}_{-96}$~~~~~~~~~~~~~~~~~~~~& \scriptsize\#(monomials) = $57$ \\ \hline
\multicolumn{2}{|c|}{\scriptsize$\tilde D_1=\tilde J_1, \tilde D_2=\tilde J_3, \tilde D_3=\tilde J_1, \tilde D_4=\tilde J_2, \tilde D_5=\tilde J_2, \tilde D_6=\tilde J_3, \tilde D_7=2\, \tilde J_1-2\, \tilde J_3+\tilde J_4, \tilde D_8=\tilde J_4$} \\ \hline
\multicolumn{2}{|c|}{\scriptsize$2\, \tilde J_1\, \tilde J_2\, \tilde J_3+4\, \tilde J_1\, \tilde J_2\, \tilde J_4+2\, \tilde J_1\, \tilde J_3\, \tilde J_4+4\, \tilde J_1\, \tilde J_4^2-8\, \tilde J_2\, \tilde J_4^2-4\, \tilde J_3\, \tilde J_4^2-16\, \tilde J_4^3$}\\ \hline \hline
$[{X_9}]^{4,28}_{-48}$ & \scriptsize\#(monomials) = $29$ \\ \hline 
\multicolumn{2}{|c|}{\scriptsize$D_1=J_3, D_2=J_1, D_3=J_1, D_4=J_2, D_5=J_2, D_6=J_3, D_7=-2\, J_1+2\, J_3+J_4, D_8=J_4$} \\ \hline
\multicolumn{2}{|c|}{\scriptsize$J_1\, J_2\, J_3+J_1\, J_3\, J_4+2\, J_2\, J_3\, J_4-2\, J_1\, J_4^2-4\, J_2\, J_4^2+2\, J_3\, J_4^2-8\, J_4^3$}\\  \hline
{\scriptsize $c_2(TX) = (12,12,12,4)$} & 
{\scriptsize
$ K= 
\left( {\begin{array}{*{20}{c}}
  1&0&0&0 \\ 
  0&0&0&1 \\ 
  0&0&1&{ - 2} \\ 
  0&1&0&0 
\end{array}} \right)$}\\ 
\hline \hline

\multicolumn{2}{c}{} \\
\multicolumn{2}{c}{Pair $10$: $\{ {\tilde x_1} \to {e^{i\pi }}{x_1},~{\tilde x_2} \to {e^{i\pi }}{x_2},~{\tilde x_3} \to {e^{i\pi }}{x_3},~{\tilde x_5} \to {e^{i\pi }}{x_5}\} $} \\
\multicolumn{2}{c}{$\{ {x_1} \to {x_2},~{x_2} \to {x_1},~{x_7} \to {x_8},~{x_8} \to {x_7}\} $} \\
\hline \hline
$~~~~~~~~~~~~~~~~~~~~[\widetilde{X_{10}}]^{4,68}_{-128}$~~~~~~~~~~~~~~~~~~~~& \scriptsize\#(monomials) = $81$ \\ \hline
\multicolumn{2}{|c|}{\scriptsize$\tilde D_1=\tilde J_1, \tilde D_2=2\, \tilde J_2+\tilde J_4, \tilde D_3=\tilde J_2, \tilde D_4=\tilde J_1, \tilde D_5=2\, \tilde J_1+\tilde J_3, \tilde D_6=\tilde J_2, \tilde D_7=\tilde J_3, \tilde D_8=\tilde J_4$} \\ \hline
\multicolumn{2}{|c|}{\scriptsize$2\, \tilde J_1\, \tilde J_2\, \tilde J_3-4\, \tilde J_2\, \tilde J_3^2+2\, \tilde J_1\, \tilde J_2\, \tilde J_4-4\, \tilde J_1\, \tilde J_4^2$}\\ \hline \hline
$[{X_{10}}]^{4,36}_{-64}$ & \scriptsize\#(monomials) = $41$ \\ \hline 
\multicolumn{2}{|c|}{\scriptsize$D_1=2\, J_2+J_3, D_2=J_1, D_3=J_2, D_4=J_1, D_5=2\, J_1+J_4, D_6=J_2, D_7=J_3, D_8=J_4$} \\ \hline
\multicolumn{2}{|c|}{\scriptsize$J_1\, J_2\, J_3-2\, J_1\, J_3^2+J_1\, J_2\, J_4-2\, J_2\, J_4^2$}\\  \hline
{\scriptsize $c_2(TX) = (12,12,0,0)$} & {\scriptsize 
$K = \left( {\begin{array}{*{20}{c}}
  0&0&1&0 \\ 
  0&1&{ - 2}&0 \\ 
  0&0&0&1 \\ 
  1&0&0&{ - 2} 
\end{array}} \right)
$}\\
\hline \hline

\multicolumn{2}{c}{} \\
\multicolumn{2}{c}{Pair $11$: $\{ {\tilde x_1} \to {e^{i\pi }}{x_1},~{\tilde x_2} \to {e^{i\pi }}{x_2},~{\tilde x_3} \to {e^{i\pi }}{x_3},~{\tilde x_4} \to {e^{i\pi }}{x_4}\} $} \\
\multicolumn{2}{c}{$\{ {x_2} \to {x_4},~{x_4} \to {x_2},~{x_5} \to {x_7},~{x_7} \to {x_5}\} $} \\
\hline \hline
$~~~~~~~~~~~~~~~~~~~~[\widetilde{X_{11}}]^{4,68}_{-128}$~~~~~~~~~~~~~~~~~~~~& \scriptsize\#(monomials) = $81$ \\ \hline
\multicolumn{2}{|c|}{\scriptsize$\tilde D_1=2\, \tilde J_1+\tilde J_4, \tilde D_2=\tilde J_3, \tilde D_3=\tilde J_2, \tilde D_4=\tilde J_1, \tilde D_5=\tilde J_1, \tilde D_6=\tilde J_2, \tilde D_7=\tilde J_3, \tilde D_8=\tilde J_4$} \\ \hline
\multicolumn{2}{|c|}{\scriptsize$2\, \tilde J_1\, \tilde J_2\, \tilde J_3+2\, \tilde J_1\, \tilde J_2\, \tilde J_4+2\, \tilde J_1\, \tilde J_3\, \tilde J_4-4\, \tilde J_2\, \tilde J_4^2-4\, \tilde J_3\, \tilde J_4^2$}\\ \hline \hline
$[{X_{11}}]^{4,36}_{-64}$ & \scriptsize\#(monomials) = $41$ \\ \hline 
\multicolumn{2}{|c|}{\scriptsize$D_1=2\, J_3+J_4, D_2=J_3, D_3=J_2, D_4=J_1, D_5=J_1, D_6=J_2, D_7=J_3, D_8=J_4$} \\ \hline
\multicolumn{2}{|c|}{\scriptsize$J_1\, J_2\, J_3+J_1\, J_3\, J_4+J_2\, J_3\, J_4-2\, J_1\, J_4^2-2\, J_2\, J_4^2$}\\ \hline
{\scriptsize $c_2(TX) = (12,12,12,0)$} & {\scriptsize 
$ K = 
\left( {\begin{array}{*{20}{c}}
  1&0&0&0 \\ 
  0&1&0&0 \\ 
  0&0&0&1 \\ 
  0&0&1&{ - 2} 
\end{array}} \right)$}\\
\hline \hline

\multicolumn{2}{c}{} \\
\multicolumn{2}{c}{Pair $12$: $\{ {\tilde x_1} \to {e^{i\pi }}{x_1},~{\tilde x_2} \to {e^{i\pi }}{x_2},~{\tilde x_3} \to {e^{i\pi }}{x_3},~{\tilde x_4} \to {e^{i\pi }}{x_4}\} $} \\
\hline \hline
$~~~~~~~~~~~~~~~~~~~~[\widetilde{X_{12}}]^{5,85}_{-160}$~~~~~~~~~~~~~~~~~~~~& \scriptsize\#(monomials) = $105$ \\ \hline
\multicolumn{2}{|c|}{\scriptsize$\tilde D_1=2\, \tilde J_1+\tilde J_3, \tilde D_2=4\, \tilde J_1+2\, \tilde J_3+\tilde J_4, \tilde D_3=\tilde J_2, \tilde D_4=\tilde J_1, \tilde D_5=\tilde J_1, \tilde D_6=\tilde J_2, \tilde D_7=\tilde J_3, \tilde D_8=\tilde J_4$} \\ \hline
\multicolumn{2}{|c|}{\scriptsize$2\, \tilde J_1\, \tilde J_2\, \tilde J_3-4\, \tilde J_2\, \tilde J_3^2+2\, \tilde J_1\, \tilde J_3\, \tilde J_4-4\, \tilde J_3^2\, \tilde J_4-4\, \tilde J_1\, \tilde J_4^2+16\, \tilde J_4^3$}\\ \hline \hline
$[{X_{12}}]^{4,44}_{-80}$ & \scriptsize\#(monomials) = $53$ \\ \hline 
\multicolumn{2}{|c|}{\scriptsize$D_1=2\, J_1+J_3,D_2=4\, J_1+2\, J_3+J_4,D_3=J_2,D_4=J_1,D_5=J_1,D_6=J_2,D_7=J_3,D_8=J_4$} \\ \hline
\multicolumn{2}{|c|}{\scriptsize$J_1\, J_2\, J_3-2\, J_2\, J_3^2+J_1\, J_3\, J_4-2\, J_3^2\, J_4-2\, J_1\, J_4^2+8\, J_4^3$}\\  \hline
{\scriptsize $ c_2(TX) = (12,12,0,-4)$} & {\scriptsize $K = \left( {\begin{array}{*{20}{c}}
  0&1&0&0 \\ 
  1&0&{ - 2}&0 \\ 
  0&0&0&1 \\ 
  0&0&1&{ - 2} 
\end{array}} \right)$ }\\
\hline \hline

\multicolumn{2}{c}{} \\
\multicolumn{2}{c}{Pair $13$: $\{ {\tilde x_1} \to {e^{i\pi }}{x_1},~{\tilde x_2} \to {e^{i\pi }}{x_2},~{\tilde x_3} \to {e^{i\pi }}{x_3},~{\tilde x_4} \to {e^{i\pi }}{x_4}\} $} \\
\multicolumn{2}{c}{$\{ {x_5} \to {x_6},~{x_6} \to {x_5}\} $}\\
\hline \hline
$~~~~~~~~~~~~~~~~~~~~[\widetilde{X_{13}}]^{5,85}_{-160}$~~~~~~~~~~~~~~~~~~~~& \scriptsize\#(monomials) = $105$ \\ \hline
\multicolumn{2}{|c|}{\scriptsize$\tilde D_1=2\, \tilde J_1+2\, \tilde J_3+\tilde J_4, \tilde D_2=\tilde J_3, \tilde D_3=\tilde J_1, \tilde D_4=\tilde J_2, \tilde D_5=\tilde J_1, \tilde D_6=\tilde J_2, \tilde D_7=\tilde J_3, \tilde D_8=\tilde J_4$} \\ \hline
\multicolumn{2}{|c|}{\scriptsize$2\, \tilde J_1\, \tilde J_2\, \tilde J_3+2\, \tilde J_1\, \tilde J_3\, \tilde J_4-4\, \tilde J_1\, \tilde J_4^2-4\, \tilde J_3\, \tilde J_4^2+16\, \tilde J_4^3$}\\ \hline \hline
$[{X_{13}}]^{4,44}_{-80}$ & \scriptsize\#(monomials) = $53$ \\ \hline 
\multicolumn{2}{|c|}{\scriptsize$D_1=2\, J_2+2\, J_3+J_4, D_2=J_3, D_3=J_2, D_4=J_1, D_5=J_1, D_6=J_2, D_7=J_3, D_8=J_4$} \\ \hline
\multicolumn{2}{|c|}{\scriptsize$J_1\, J_2\, J_3+J_2\, J_3\, J_4-2\, J_2\, J_4^2-2\, J_3\, J_4^2+8\, J_4^3$}\\  \hline
{\scriptsize $c_2(TX) = (12,12,12,-4)$} & 
{\scriptsize $ K = 
\left( {\begin{array}{*{20}{c}}
  1&0&0&0 \\ 
  0&0&0&1 \\ 
  0&1&0&{ - 2} \\ 
  0&0&1&{ - 2} 
\end{array}} \right)
$}
\\
\hline \hline

\multicolumn{2}{c}{} \\
\multicolumn{2}{c}{Pair $14$: $\{ {\tilde x_1} \to {e^{i\pi }}{x_1},~{\tilde x_2} \to {e^{i\pi }}{x_2},~{\tilde x_3} \to {e^{i\pi }}{x_3},~{\tilde x_4} \to {e^{i\pi }}{x_4}\} $} \\
\hline \hline
$~~~~~~~~~~~~~~~~~~~~[\widetilde{X_{14}}]^{3,115}_{-224}$~~~~~~~~~~~~~~~~~~~~& \scriptsize\#(monomials) = $153$ \\ \hline
\multicolumn{2}{|c|}{\scriptsize$\tilde D_1=2\, \tilde J_1+2\, \tilde J_2+2\, \tilde J_3, \tilde D_2=\tilde J_3, \tilde D_3=\tilde J_2, \tilde D_4=\tilde J_1, \tilde D_5=\tilde J_1, \tilde D_6=\tilde J_2, \tilde D_7=\tilde J_3$} \\ \hline
\multicolumn{2}{|c|}{\scriptsize$2\, \tilde J_1\, \tilde J_2\, \tilde J_3$}\\ \hline \hline
$[{X_{14}}]^{3,59}_{-112}$ & \scriptsize\#(monomials) = $77$ \\ \hline 
\multicolumn{2}{|c|}{\scriptsize$D_1=2\, J_1+2\, J_2+2\, J_3, D_2=J_3, D_3=J_2, D_4=J_1, D_5=J_1, D_6=J_2, D_7=J_3$} \\ \hline
\multicolumn{2}{|c|}{\scriptsize$ J_1\, J_2\, J_3$}\\ \hline
\multicolumn{2}{|c|}{\scriptsize $c_2(TX) = (12,12,12)$}\\ \hline
\multicolumn{2}{|c|}{\scriptsize 
$K_1 = \left( {\begin{array}{*{20}{c}}
  0&1&0 \\ 
  1&-1&0 \\ 
  0&-1&1 
\end{array}} \right)$ ,
$K_2 = \left( {\begin{array}{*{20}{c}}
  0&0&1 \\ 
  0&1&{ -1} \\ 
  1&0&{ - 1} 
\end{array}} \right)$,
$K_3 = \left( {\begin{array}{*{20}{c}}
  1&0&0 \\ 
  -1&0&1 \\ 
  -1&1&0 
\end{array}} \right)$,
$K_{join} = \left( {\begin{array}{*{20}{c}}
  1&0&0 \\ 
  0&1&0 \\ 
  0&0&1 
\end{array}} \right)$
}\\
\hline \hline

\caption{\em Summary of the Calabi-Yau three-fold geometries, for both upstairs manifolds $\widetilde{X_i}$ and downstairs manifolds $X_i$. The phase rotation rule (together with the permutation if needed) is specified at the start of each geometry pair. The Hodge numbers $h^{1,1}$ and $h^{2,1}$, as well as the Euler Character $\chi$ of the manifold $X$ are presented as $[X_i]^{h^{1,1}, h^{2,1}}_\chi$. Further geometrical properties follow in order: number of generating monomials, Picard group structure and intersection polynomial, as well as $c_2(TX_i)$ and K\"ahler cone matrix for the downstairs spaces.}
\label{BaseGeometries}
\end{longtable}}

\section{GUT Models} \label{su5result}

\setlongtables 
\LTcapwidth=\textwidth
{\renewcommand{\arraystretch}{1.07}
\begin{longtable}[thb]{|c|c|}
\hline
\endhead
\hline
\multicolumn{2}{r}{\scriptsize \slshape continued in the next page}
\endfoot
\endlastfoot
\hline 
\multicolumn{2}{c}{\scriptsize  Downstairs Rank-5 GUT Models} \\ \hline \hline
\endfirsthead 
\hline 
\multicolumn{2}{c}{\scriptsize  Downstairs Rank-5 GUT Models} \\ \hline \hline
\endhead 
\small$\left[X_3\right]^{4, 36}_{-64}$ & $\pi_1(X_3)=\mathbb Z_2$ \\ \hline
\tiny\{(-1, 2, 2, 0),(0, -1, 1, 0),(0, -1, 1, 0),(0, 0, -3, 1),(1, 0, -1, -1)\} 
&\tiny \{(-1, 1, 3, 0),(0, 1, -1, 0),(0, 1, -1, 0),(0, 1, -1, 0),(1, -4, 0, 0)\}
\\ \hline 
\tiny\{(-1, 1, 3, 0),(0, 1, -1, 0),(0, 1, -1, 0),(0, -4, 0, 1),(1, 1, -1, -1)\}
&\tiny \{(-1, 1, 2, 0),(0, 1, -1, 0),(0, 1, -1, 0),(0, -4, 0, 1),(1, 1, 0, -1)\}
\\ \hline
\tiny \{(-1, 1, 1, 0),(0, 1, 1, -2),(0, -1, 0, 1),(0, -1, 0, 1),(1, 0, -2, 0)\}
&\tiny \{(-1, 0, 1, 0),(-1, 1, 0, -1),(-1, 0, 1, 0),(1, 0, 0, -1),(2, -1, -2, 2)\}
\\ \hline
\tiny \{(-1, 0, 1, 0),(-1, 0, 1, 0),(-1, 0, 1, 0),(1, 1, -1, -2),(2, -1, -2, 2)\}
& \tiny\{(-1, 0, 1, 0),(-1, 1, -1, 0),(0, 1, 2, -2),(1, -1, -1, 1),(1, -1, -1, 1)\}
\\ \hline
\tiny\{(-1, 1, 1, -1),(-2, -1, 1, 1),(-1, 1, 1, -1),(2, 1, -2, 0),(2, -2, -1, 1)\}
& \tiny\{(-1, 1, 1, -1),(-1, 1, 1, -1),(-1, 1, 1, -1),(1, -3, -1, 2),(2, 0, -2, 1)\}
\\ \hline \hline
\small$\left[X_6\right]^{3, 59}_{-112}$ & $\pi_1(X_6)=\mathbb Z_2$ \\ \hline
\tiny\{(-3, 0, 1), (0, 3, -1), (1, -1, 0), (1, -1, 0), (1, -1, 0)\}
&\tiny \{(-1, 1, 0), (-1, 1, 0), (-1, 1, 0), (1, -4, 1), (2, 1, -1)\}
\\ \hline \hline
\small$\left[X_9\right]^{4, 28}_{-48}$ & $\pi_1(X_9)=\mathbb Z_2$ \\ \hline
\tiny\{(-4, 0, 1, 1),(1, 3, -1, -1),(1, -1, 0, 0),(1, -1, 0, 0),(1, -1, 0, 0)\}
&\tiny \{(-3, 1, -1, 1),(0, 2, 1, -1),(1, -1, 0, 0),(1, -1, 0, 0),(1, -1, 0, 0)\}
\\ \hline
\tiny\{(-3, 1, 0, 1),(0, 2, 0, -1),(1, -1, 0, 0),(1, -1, 0, 0),(1, -1, 0, 0)\}
&\tiny\{(-2, 3, 0, -1),(-1, 0, 0, 1),(1, -1, 0, 0),(1, -1, 0, 0),(1, -1, 0, 0)\}
\\ \hline
\tiny\{(-2, 1, 1, 0),(-1, -2, 2, 1),(1, 1, -1, -1),(1, 0, -1, 0),(1, 0, -1, 0)\} 
& \tiny\{(-2, 0, 0, 1),(-1, 3, 0, -1),(1, -1, 0, 0),(1, -1, 0, 0),(1, -1, 0, 0)\} 
\\ \hline
\tiny\{(-2, 1, 0, 1),(-1, 2, 0, -1),(1, -1, 0, 0),(1, -1, 0, 0),(1, -1, 0, 0)\}
&\tiny \{(-2, 0, 1, 2),(-1, 3, -1, -2),(1, -1, 0, 0),(1, -1, 0, 0),(1, -1, 0, 0)\}
\\ \hline
\tiny\{(-1, 1, 0, 0),(-1, 1, 0, 0),(-1, 1, 0, 0),(-1, -1, -1, 1),(4, -2, 1, -1)\}
&\tiny \{(-1, 1, 0, 0),(-1, 1, 0, 0),(-1, 1, 0, 0),(0, -4, 1, 1),(3, 1, -1, -1)\}
\\ \hline
\tiny\{(-1, 1, 3, 0),(0, 1, -1, 0),(0, 1, -1, 0),(0, 1, -1, 0),(1, -4, 0, 0)\}
&\tiny\{(-1, -4, 2, 1),(0, 1, -1, 0),(0, 1, -1, 0),(0, 1, -1, 0),(1, 1, 1, -1)\}
\\ \hline \hline
\small$\left[X_{10}\right]^{4, 36}_{-64}$ & $\pi_1(X_{10})=\mathbb Z_2$ \\ \hline
\tiny\{(-3, 4, 2, -1),(1, -1, -1, 0),(1, -1, -1, 0),(1, -1, -1, 0),(0, -1, 1, 1)\} 
&\tiny \{(-3, 4, 2, -2),(1, -1, -1, 1),(1, -1, -1, 1),(1, -1, -1, 1),(0, -1, 1, -1)\}
\\ \hline
\tiny\{(-4, 3, 2, -1),(2, -2, -1, 1),(2, -2, -1, 1),(2, -2, -1, 1),(-2, 3, 1, -2)\} 
&\tiny \{(-2, 1, 2, 1),(5, -4, -2, 2),(-1, 1, 0, -1),(-1, 1, 0, -1),(-1, 1, 0, -1)\}
\\ \hline
\tiny\{(0, 1, 2, -2),(1, -1, -1, 1),(1, -1, -1, 1),(1, -1, -1, 1),(-3, 2, 1, -1)\} 
&\tiny \{(-1, 1, 1, 0),(1, 0, -2, -1),(2, -3, -1, 1),(-1, 1, 1, 0),(-1, 1, 1, 0)\}
\\ \hline
\tiny\{(-3, 1, 1, 0),(0, -1, -1, 1),(1, -1, 0, 0),(1, -1, 0, 0),(1, 2, 0, -1)\}
&\tiny \{(-4, 1, 1, 0),(2, -2, -1, 1),(1, -1, 0, 0),(1, -1, 0, 0),(0, 3, 0, -1)\}
\\ \hline
\tiny\{(-4, 1, 1, 0),(2, -1, -1, 1),(1, -1, 0, 0),(1, -1, 0, 0),(0, 2, 0, -1)\} 
&\tiny\{(-4, 1, 1, 0),(1, 2, -1, 0),(1, -1, 0, 0),(1, -1, 0, 0),(1, -1, 0, 0)\} 
\\ \hline
\tiny\{(-2, 1, 1, -1),(2, 1, -1, 0),(1, -1, 0, 0),(1, -1, 0, 0),(-2, 0, 0, 1)\} 
&\tiny \{(-1, 0, 1, 1),(2, -1, -2, 0),(3, -3, -1, 1),(-2, 2, 1, -1),(-2, 2, 1, -1)\} 
\\ \hline
\tiny\{(-1, 0, 1, 0),(0, 0, -1, 1),(0, 0, -1, 1),(1, -1, 0, 0),(0, 1, 1, -2)\} 
&\tiny\{(0, -1, 1, 1),(1, 0, -2, 0),(-1, 1, -1, 1),(0, 0, 1, -1),(0, 0, 1, -1)\} 
\\ \hline
\tiny\{(0, -3, 1, 0),(3, 0, -1, 0),(-1, 1, 0, 0),(-1, 1, 0, 0),(-1, 1, 0, 0)\} 
&\tiny \{(0, -3, 1, 0),(1, 0, -1, 1),(1, 1, 0, -1),(-1, 1, 0, 0),(-1, 1, 0, 0)\} 
\\ \hline
\tiny\{(-1, 2, 0, 1),(1, -2, -3, 2),(0, 0, 1, -1),(0, 0, 1, -1),(0, 0, 1, -1)\} 
&\tiny\{(-1, 1, 0, 0),(1, 0, -2, 1),(0, -1, 0, 1),(0, 0, 1, -1),(0, 0, 1, -1)\} 
\\ \hline
\tiny\{(-1, 1, 0, 0),(2, 1, -1, 0),(1, -3, 0, 1),(-1, 1, 0, 0),(-1, 0, 1, -1)\} 
&\tiny\{(-3, 0, 0, 1),(1, 1, -1, 0),(1, -1, 0, 0),(1, -1, 0, 0),(0, 1, 1, -1)\} 
\\ \hline
\tiny\{(-3, 0, 0, 1),(1, 2, -1, 0),(1, -1, 0, 0),(1, -1, 0, 0),(0, 0, 1, -1)\} 
&\tiny \{(0, -1, 0, 1),(3, -2, -2, 1),(1, -1, 0, 0),(-2, 2, 1, -1),(-2, 2, 1, -1)\} 
\\ \hline
\tiny\{(-1, 0, -1, 2),(1, 0, -2, 1),(0, 0, 1, -1),(0, 0, 1, -1),(0, 0, 1, -1)\} 
&\tiny \{(-1, 0, -1, 1),(4, -3, -2, 2),(-1, 1, 1, -1),(-1, 1, 1, -1),(-1, 1, 1, -1)\} 
\\ \hline
\tiny\{(1, 0, -2, 2),(2, -3, -1, 1),(-1, 1, 1, -1),(-1, 1, 1, -1),(-1, 1, 1, -1)\}
&
\\ \hline \hline
\small$\left[X_{11}\right]^{4, 36}_{-64}$ & $\pi_1(X_{11})=\mathbb Z_2$ \\ \hline
\tiny\{(2, 2, -1, -1), (-3, 0, 1, 0), (-1, 0, 0, 1), (1, -1, 0, 0), (1, -1, 0, 0)\}
&
\tiny\{(2, 2, -3, -1), (-2, 1, 0, 1), (0, -1, 1, 0), (0, -1, 1, 0), (0, -1, 1, 0)\}
\\ \hline
\tiny\{(2, 2, -3, -1), (-1, -1, 1, 1), (-1, 0, 1, 0), (-1, 0, 1, 0), (1, -1, 0, 0)\}
&
\tiny\{(2, 2, -3, -2), (-2, 1, 0, -1), (0, -1, 1, 1), (0, -1, 1, 1), (0, -1, 1, 1)\}
\\ \hline
\tiny\{(1, 3, -1, 0), (-4, 0, 1, 0), (1, -1, 0, 0), (1, -1, 0, 0), (1, -1, 0, 0)\}
&
\tiny\{(1, 3, -1, -1), (-4, 0, 1, 1), (1, -1, 0, 0), (1, -1, 0, 0), (1, -1, 0, 0)\}
\\ \hline
\tiny\{(1, 2, -1, 0), (-4, 0, 1, 1), (1, -1, 0, 0), (1, -1, 0, 0), (1, 0, 0, -1)\}
&
\tiny\{(1, 2, -3, -1), (-1, -1, 2, 1), (-1, -1, 2, 1), (0, 2, -1, -2), (1, -2, 0, 1)\}
\\ \hline
\tiny\{(1, 2, -4, -1), (-1, -1, 2, 1), (-1, -1, 2, 1), (-1, -1, 2, 1), (2, 1, -2, -2)\}
&
\tiny\{(1, 2, -2, -2), (-1, -1, 2, 1), (-1, -1, 2, 1), (0, 1, -1, 0), (1, -1, -1, 0)\}
\\ \hline
\tiny\{(1, 1, -2, 0), (-2, 0, 1, 1), (0, -1, 1, 0), (0, -1, 1, 0), (1, 1, -1, -1)\}
&
\tiny\{(1, 1, -1, -1), (-4, 0, 1, 1), (1, -1, 2, 0), (1, 0, -1, 0), (1, 0, -1, 0)\}
\\ \hline
\tiny\{(1, 1, -2, -1), (-4, 0, 1, 1), (1, -1, 3, 0), (1, 0, -1, 0), (1, 0, -1, 0)\}
&
\tiny\{(1, 1, -2, -1), (-3, -1, 3, 1), (0, -2, 3, 2), (1, 1, -2, -1), (1, 1, -2, -1)\}
\\ \hline
\tiny\{(1, 1, -2, -1), (-3, -1, 3, 2), (0, -2, 3, 1), (1, 1, -2, -1), (1, 1, -2, -1)\}
&
\tiny\{(1, 1, -2, -1), (-2, -1, 3, 1), (-1, 1, -1, 1), (1, -2, 2, 0), (1, 1, -2, -1)\}
\\ \hline
\tiny\{(1, 1, -2, -1), (-2, 0, 3, 1), (0, -1, 1, 0), (0, -1, 0, 1), (1, 1, -2, -1)\}
&
\tiny\{(1, 1, -3, -1), (-1, 0, 1, 1), (-1, 0, 1, 1), (-1, 0, 1, 1), (2, -1, 0, -2)\}
\\ \hline
\tiny\{(1, 1, -3, -2), (-1, 0, 1, 0), (-1, 0, 1, 0), (-1, 0, 1, 0), (2, -1, 0, 2)\}
&
\tiny\{(0, 3, 0, -1), (-1, -1, 2, 1), (0, -1, 1, 0), (0, -1, 1, 0), (1, 0, -4, 0)\}
\\ \hline
\tiny\{(0, 3, 0, -1), (0, -1, 1, 0), (0, -1, 1, 0), (0, -1, 1, 0), (0, 0, -3, 1)\}
&
\tiny\{(0, 2, -1, 1), (-3, 1, 1, 2), (1, -1, 0, -1), (1, -1, 0, -1), (1, -1, 0, -1)\}
\\ \hline
\tiny\{(0, 2, 1, -1), (-1, -1, 1, 1), (0, 1, -1, 0), (0, 1, -1, 0), (1, -3, 0, 0)\}
&
\tiny\{(0, 2, 1, -1), (-1, 0, 0, 1), (0, -1, 1, 0), (0, -1, 1, 0), (1, 0, -3, 0)\}
\\ \hline
\tiny\{(0, 2, 0, -1), (-1, 0, 2, 1), (0, -1, 1, 0), (0, -1, 1, 0), (1, 0, -4, 0)\}
&
\tiny\{(0, 2, -1, -2), (-3, 1, 1, -1), (1, -1, 0, 1), (1, -1, 0, 1), (1, -1, 0, 1)\}
\\ \hline
\tiny\{(0, 1, -1, 0), (-2, -1, 2, 1), (0, 1, -1, 0), (0, 1, -1, 0), (2, -2, 1, -1)\}
&
\tiny\{(0, 1, -1, 0), (-2, 0, 1, 1), (0, 1, -1, 0), (1, -2, 1, 0), (1, 0, 0, -1)\}
\\ \hline
\tiny\{(0, 1, -1, 0), (-2, 2, 1, -1), (0, -1, 0, 1), (1, -1, 0, 0), (1, -1, 0, 0)\}
&
\tiny\{(0, 1, -1, 0), (-2, 2, 1, 1), (0, -1, 0, 1), (1, -1, 0, -1), (1, -1, 0, -1)\}
\\ \hline
\tiny\{(0, 1, -1, 0), (-1, -2, 4, 2), (0, 1, -1, 0), (0, 1, -1, 0), (1, -1, -1, -2)\}
&
\tiny\{(0, 1, -1, 0), (-1, 0, 1, 0), (-1, 0, 0, 1), (0, 1, -1, 0), (2, -2, 1, -1)\}
\\ \hline
\tiny\{(0, 1, -1, 0), (-1, 1, 3, 0), (0, 1, -1, 0), (0, 1, -1, 0), (1, -4, 0, 0)\}
&
\tiny\{(0, 1, -1, 0), (-1, 1, 0, 1), (0, 1, 2, -1), (0, 1, -1, 0), (1, -4, 0, 0)\}
\\ \hline
\tiny\{(0, 1, -1, 0), (-1, 1, 1, 1), (0, 1, 1, -1), (0, 1, -1, 0), (1, -4, 0, 0)\}
&
\tiny\{(0, 1, -1, 0), (-1, 2, 2, 0), (0, -3, 1, 1), (0, 1, -1, 0), (1, -1, -1, -1)\}
\\ \hline
\tiny\{(0, 1, -1, 0), (0, -4, 1, 1), (0, 1, 2, -1), (0, 1, -1, 0), (0, 1, -1, 0)\}
&
\tiny\{(0, 1, -4, 0), (-1, 0, 1, 0), (-1, 0, 1, 0), (-1, 0, 1, 0), (3, -1, 1, 0)\}
\\ \hline
\tiny\{(0, 1, 0, -1), (-2, 1, 1, 0), (0, 0, -1, 1), (1, -1, 0, 0), (1, -1, 0, 0)\}
&
\tiny\{(0, 1, 0, -1), (-1, 0, 1, 0), (-1, 0, 1, 0), (0, 0, -3, 1), (2, -1, 1, 0)\}
\\ \hline
\tiny\{(0, 1, -1, -1), (-3, -1, 3, 1), (0, 1, -1, -1), (0, 1, -1, -1), (3, -2, 0, 2)\}
&
\tiny\{(0, 1, -1, -1), (-2, -2, 5, 2), (0, 1, -1, -1), (0, 1, -1, -1), (2, -1, -2, 1)\}
\\ \hline
\tiny\{(0, 1, -1, -1), (-2, -1, 2, 1), (0, 1, -1, -1), (0, 1, -1, -1), (2, -2, 1, 2)\}
&
\tiny\{(0, 1, -1, -1), (-1, -2, 4, 2), (0, 1, -1, -1), (0, 1, -1, -1), (1, -1, -1, 1)\}
\\ \hline
\tiny\{(0, 1, -2, -1), (-1, 0, 1, 0), (-1, 0, 1, 0), (0, 0, -2, 1), (2, -1, 2, 0)\}
&
\tiny\{(0, 1, 0, -2), (-3, 2, 0, -1), (1, -1, 0, 1), (1, -1, 0, 1), (1, -1, 0, 1)\}
\\ \hline
\tiny\{(0, 1, 0, -2), (-1, 0, 1, 0), (-1, 1, -1, 0), (1, -1, 0, 1), (1, -1, 0, 1)\}
&
\tiny\{(0, 1, 0, -2), (-1, 1, 0, 1), (-1, 1, 0, 1), (1, -2, 1, -1), (1, -1, -1, 1)\}
\\ \hline
\tiny\{(0, 0, -3, 1), (-1, 0, 1, 0), (-1, 0, 1, 0), (-1, 1, 0, -1), (3, -1, 1, 0)\}
&
\tiny\{(-1, 2, -1, 1), (-2, 2, 1, 0), (1, -2, 0, 1), (1, -1, 0, -1), (1, -1, 0, -1)\}
\\ \hline
\tiny\{(-1, 2, -2, 1), (-2, 1, 2, 2), (1, -1, 0, -1), (1, -1, 0, -1), (1, -1, 0, -1)\}
&
\tiny\{(-1, 2, -2, -3), (-2, 1, 2, 0), (1, -1, 0, 1), (1, -1, 0, 1), (1, -1, 0, 1)\}
\\ \hline
\tiny\{(-1, 1, -1, 2), (-2, 2, 1, 1), (1, -1, 0, -1), (1, -1, 0, -1), (1, -1, 0, -1)\}
&
\tiny\{(-1, 1, -1, 1), (-2, 2, 1, -1), (1, -1, 0, 0), (1, -1, 0, 0), (1, -1, 0, 0)\}
\\ \hline \hline 
\small$\left[X_{12}\right]^{4, 44}_{-80}$ & $\pi_1(X_{12})=\mathbb Z_2$ \\ \hline
\tiny \{(1, -4, 1, 0),(-1, 1, 0, 0),(-1, 1, 0, 0),(-1, 1, 0, 0),(2, 1, -1, 0)\}
&
\\ \hline \hline
\small$\left[X_{13}\right]^{4, 44}_{-80}$ & $\pi_1(X_{13})=\mathbb Z_2$ \\ \hline
\tiny\{(3, 1, -1, 0),(-1, 1, 0, 0),(-1, 1, 0, 0),(-1, 1, 0, 0),(0, -4, 1, 0)\}
&
\tiny\{(3, -1, 1, 0),(-1, 0, 1, 0),(-1, 0, 1, 0),(-1, 0, 1, 0),(0, 1, -4, 0)\}
\\ \hline
\tiny\{(3, 1, -1, -1),(-1, 1, 0, 0),(-1, 1, 0, 0),(-1, 1, 0, 0),(0, -4, 1, 1)\}
&
\tiny\{(3, -1, 1, -1),(-1, 0, 1, 0),(-1, 0, 1, 0),(-1, 0, 1, 0),(0, 1, -4, 1)\} 
\\ \hline
\tiny\{(2, 1, -4, -1),(-1, 0, 1, 0),(-1, 0, 1, 0),(-1, 0, 1, 0),(1, -1, 1, 1)\} 
&
\tiny\{(2, -4, 1, -1),(-1, 1, 0, 0),(-1, 1, 0, 0),(-1, 1, 0, 0),(1, 1, -1, 1)\} 
\\ \hline 
\tiny\{(1, 3, -1, 0),(-4, 0, 1, 0),(1, -1, 0, 0),(1, -1, 0, 0),(1, -1, 0, 0)\} 
&
\tiny\{(1, 0, -1, 0),(-4, 1, 0, 0),(1, -1, 3, 0),(1, 0, -1, 0),(1, 0, -1, 0)\} 
\\ \hline 
\tiny\{(1, 0, -1, 0),(-4, 1, 0, 1),(1, -1, 3, -1),(1, 0, -1, 0),(1, 0, -1, 0)\} 
&
\tiny\{(1, 0, -1, 0),(-2, 1, 0, -1),(-1, -1, 3, 1),(1, 0, -1, 0),(1, 0, -1, 0)\} 
\\ \hline 
\tiny\{(1, 0, -3, 0),(-1, 1, 1, 1),(0, 1, 0, -1),(0, -1, 1, 0),(0, -1, 1, 0)\} 
&
\tiny\{(1, 0, -4, 0),(-1, 3, 1, 0),(0, -1, 1, 0),(0, -1, 1, 0),(0, -1, 1, 0)\} 
\\ \hline 
\tiny\{(1, -1, 0, 0),(-4, 0, 1, 1),(1, 3, -1, -1),(1, -1, 0, 0),(1, -1, 0, 0)\} 
&
\tiny\{(1, -1, 0, 0),(-3, 5, -1, 2),(0, -2, 1, -2),(1, -1, 0, 0),(1, -1, 0, 0)\} 
\\ \hline 
\tiny\{(1, -1, 0, 0),(-2, 0, 1, -1),(-1, 3, -1, 1),(1, -1, 0, 0),(1, -1, 0, 0)\} 
&
\tiny\{(1, -3, 0, 0),(-1, 1, 1, 1),(0, 0, 1, -1),(0, 1, -1, 0),(0, 1, -1, 0)\}
\\ \hline 
\tiny\{(1, -4, 0, 0),(-1, 1, 3, 0),(0, 1, -1, 0),(0, 1, -1, 0),(0, 1, -1, 0)\}
&
\\ \hline \hline
\small$\left[X_{14}\right]^{3, 59}_{-112}$ & $\pi_1(X_{14})=\mathbb Z_2$ \\ \hline
\tiny \{(-1, 1, 3), (0, 1, -1), (0, 1, -1), (0, 1, -1), (1, -4, 0)\}
&
\\ \hline
\caption{\em Heterotic $SU(5)$-GUT models on the downstairs Calabi-Yau three-folds $[X_i]_{\chi}^{h^{1,1}, h^{2,1}}$ with $\pi_1 \neq \phi$. The superscripts and the subscript denote, respectively, Hodge numbers and Euler character of the Calabi-Yau base. The gauge bundle of each model is a Whitney sum of five line bundles.}
\label{mainresult}
\end{longtable}}

\setlongtables 
\LTcapwidth=\textwidth
{\renewcommand{\arraystretch}{1.07}
\begin{longtable}[thb]{|c|c|}
\hline
\endhead
\hline
\multicolumn{2}{r}{\scriptsize \slshape continued in the next page}
\endfoot
\endlastfoot
\hline 
\multicolumn{2}{c}{\scriptsize  Downstairs Rank-4 GUT Models} \\ \hline \hline
\endfirsthead 
\hline 
\multicolumn{2}{c}{\scriptsize  Downstairs Rank-4 GUT Models} \\ \hline \hline
\endhead 
\small$\left[X_5\right]^{3, 43}_{-80}$ & $\pi_1(X_5)=\mathbb Z_2$ \\ \hline
\tiny \{(3, 3, -1), (-2, 2, 0), (1, -1, 0), (-2, -4, 1)\}
&
\tiny \{(3, 3, -1), (1, -1, 0), (2, -2, 0), (-6, 0, 1)\}
\\ \hline
\tiny \{(5, 1, -1), (-2, 2, 0), (-1, 1, 0), (-2, -4, 1)\}
&
\tiny \{(5, 1, -1), (-2, 2, 0), (3, -3, 0), (-6, 0, 1)\}
\\ \hline
\tiny \{(5, 1, -1), (-1, 1, 0), (2, -2, 0), (-6, 0, 1\}
&
\\ \hline \hline
\small$\left[X_6\right]^{3, 59}_{-112}$ & $\pi_1(X_6)=\mathbb Z_2$ \\ \hline
\tiny \{(2, 1, -1), (-1, 1, 0), (-2, 2, 0), (1, -4, 1)\}
&
\tiny \{(6, 1, -1), (-2, 1, 0), (-4, 2, 0), (0, -4, 1)\}
\\ \hline
\tiny \{(3, 2, -1), (3, -3, 0), (-2, 2, 0), (-4, -1, 1)\}
&
\tiny \{(3, 2, -1), (2, -2, 0), (-1, 1, 0), (-4, -1, 1)\}
\\ \hline
\tiny \{(3, 2, -1), (1, -1, 0), (-4, 4, 0), (0, -5, 1)\}
&
\tiny \{(3, 2, -1), (-1, 1, 0), (-2, 2, 0), (0, -5, 1)\}
\\ \hline
\tiny \{(0, 3, -1), (2, -2, 0), (1, -1, 0), (-3, 0, 1)\}
&
\tiny \{(0, 3, -1), (1, -1, 0), (-2, 2, 0), (1, -4, 1)\}
\\ \hline
\tiny \{(1, 4, -1), (4, -4, 0), (-1, 1, 0), (-4, -1, 1)\}
&
\tiny \{(1, 4, -1), (2, -2, 0), (1, -1, 0), (-4, -1, 1)\}
\\ \hline
\tiny \{(1, 4, -1), (2, -2, 0), (-3, 3, 0), (0, -5, 1)\}
&
\tiny \{(1, 4, -1), (1, -1, 0), (-2, 2, 0), (0, -5, 1)\}
\\ \hline
\tiny \{(0, 7, -1), (2, -4, 0), (1, -2, 0), (-3, -1, 1\}
&
\\ \hline \hline
\small$\left[X_8\right]^{3, 75}_{-144}$ & $\pi_1(X_8)=\mathbb Z_2$ \\ \hline
\tiny \{(1, -3, 1), (2, -2, 0), (-3, 3, 0), (0, 2, -1)\}
&
\tiny \{(1, -3, 1), (1, -1, 0), (-2, 2, 0), (0, 2, -1)\}
\\ \hline
\tiny \{(1, -3, 1), (1, -1, 0), (2, 0, -1), (-4, 4, 0)\}
&
\tiny \{(1, -3, 1), (-1, 1, 0), (-2, 2, 0), (2, 0, -1)\}
\\ \hline
\tiny \{(4, -4, 0), (-3, 1, 1), (-1, 1, 0), (0, 2, -1)\}
&
\tiny \{(4, -4, 0), (-3, 1, 1), (2, 0, -1), (-3, 3, 0)\}
\\ \hline
\tiny \{(3, -3, 0), (-3, 1, 1), (-2, 2, 0), (2, 0, -1)\}
&
\tiny \{(2, -2, 0), (-3, 1, 1), (1, -1, 0), (0, 2, -1)\}
\\ \hline
\tiny \{(2, -2, 0), (-3, 1, 1), (-1, 1, 0), (2, 0, -1\}
&
\\ \hline \hline
\small$\left[X_{14}\right]^{3, 59}_{-112}$ & $\pi_1(X_{14})=\mathbb Z_2$ \\ \hline
\tiny \{(1, -1, -5), (0, 2, -2), (0, -3, 3), (-1, 2, 4)\}
&
\tiny \{(1, -1, -5), (0, 1, -1), (0, -2, 2), (-1, 2, 4)\}
\\ \hline
\tiny \{(1, -1, -5), (0, 1, -1), (-1, 4, 2), (0, -4, 4)\}
&
\tiny \{(1, -1, -5), (0, -1, 1), (0, -2, 2), (-1, 4, 2)\}
\\ \hline
\tiny \{(-1, 1, -5), (2, 0, -2), (-3, 0, 3), (2, -1, 4)\}
&
\tiny \{(-1, 1, -5), (1, 0, -1), (4, -1, 2), (-4, 0, 4)\}
\\ \hline
\tiny \{(-1, 1, -5), (1, 0, -1), (-2, 0, 2), (2, -1, 4)\}
&
\tiny \{(-1, 1, -5), (-1, 0, 1), (4, -1, 2), (-2, 0, 2)\}
\\ \hline
\tiny \{(1, -1, -4), (0, -2, 1), (-1, 7, 1), (0, -4, 2)\}
&
\tiny \{(1, 0, -4), (0, 2, -2), (0, -3, 3), (-1, 1, 3)\}
\\ \hline
\tiny \{(1, 0, -4), (0, 1, -1), (0, -2, 2), (-1, 1, 3)\}
&
\tiny \{(1, 0, -4), (0, -1, 1), (-1, 3, 1), (0, -2, 2)\}
\\ \hline
\tiny \{(2, 0, -4), (1, 0, -2), (-4, 1, -1), (1, -1, 7)\}
&
\tiny \{(4, 0, -4), (-5, 1, -1), (-1, 0, 1), (2, -1, 4)\}
\\ \hline
\tiny \{(-1, 1, -4), (7, -1, 1), (-2, 0, 1), (-4, 0, 2)\}
&
\tiny \{(0, 1, -4), (2, 0, -2), (1, -1, 3), (-3, 0, 3)\}
\\ \hline
\tiny \{(0, 1, -4), (1, 0, -1), (-2, 0, 2), (1, -1, 3)\}
&
\tiny \{(0, 1, -4), (3, -1, 1), (-1, 0, 1), (-2, 0, 2)\}
\\ \hline
\tiny \{(0, 2, -4), (0, 1, -2), (1, -4, -1), (-1, 1, 7)\}
&
\tiny \{(0, 4, -4), (1, -5, -1), (0, -1, 1), (-1, 2, 4)\}
\\ \hline
\tiny \{(3, 0, -3), (-5, 1, -1), (4, -1, 2), (-2, 0, 2)\}
&
\tiny \{(3, 0, -3), (-4, 1, 0), (3, -1, 1), (-2, 0, 2)\}
\\ \hline
\tiny \{(0, 3, -3), (1, -5, -1), (0, -2, 2), (-1, 4, 2)\}
&
\tiny \{(0, 3, -3), (1, -4, 0), (-1, 3, 1), (0, -2, 2)\}
\\ \hline
\tiny \{(2, 0, -2), (1, 0, -1), (-5, 1, -1), (2, -1, 4)\}
&
\tiny \{(2, 0, -2), (1, 0, -1), (-4, 1, 0), (1, -1, 3)\}
\\ \hline
\tiny \{(2, 0, -2), (-5, 1, -1), (-1, 0, 1), (4, -1, 2)\}
&
\tiny \{(2, 0, -2), (-4, 1, 0), (3, -1, 1), (-1, 0, 1\}
\\ \hline \hline
\caption{\em Heterotic $SO(10)$-GUT models on the Calabi-Yau three-folds $[X_i]_{\chi}^{h^{1,1}, h^{2,1}}$ with $h^{1,1}=3$ and $\pi_1 \neq \phi$. The superscripts and the subscript denote, respectively, Hodge numbers and Euler character of the Calabi-Yau base. The gauge bundle of each model is a Whitney sum of four line bundles.}
\label{mainresult2}
\end{longtable}}

\section*{Acknowledgments}
A.~L.~is partially supported by the EC 6th Framework Programme MRTN-CT-2004-503369 and by the EPSRC network grant EP/l02784X/1. YHH would like to thank the Science and
Technology Facilities Council, UK, for grant ST/J00037X/1,
the Chinese Ministry of Education, for a Chang-Jiang Chair Professorship at NanKai University as well as the City of Tian-Jin for a Qian-Ren Scholarlship, the US NSF for grant CCF-1048082, as well as City University, London and Merton College, Oxford, for their enduring support. Chuang Sun would like to thank Andrei Constantin for helpful discussion. S.-J.~L. thanks the University of Oxford for hospitality while part of this work was completed.

~\\
~\\
~\\

\end{document}